\documentclass[pre,reprint,graphicx,twoside]{revtex4-1}

\usepackage{graphicx,epstopdf,amsfonts,amsmath,amssymb,array,mathrsfs}
\usepackage{enumerate}

\usepackage{fancyhdr}

\def\be{\begin{equation}}
\def\ee{\end{equation}}

\def\cvp{\raise 2pt\hbox{,}}  \def\cvd{\raise 2pt\hbox{.}}

\def\d{{\rm d}}
\def\trans{{\mathsf T}}
\def\br{\boldsymbol r} 
\def\bm{\boldsymbol \mu} \def\bmm{\boldsymbol m} 
\def\bL{\boldsymbol \ell} \def\bW{\boldsymbol w}

\def\gcs{\textsc{gcs}}

\begin{document}



\title{Investigating complex networks with inverse models: \\
Analytical aspects of spatial leakage and connectivity estimation} 

\author{Vincent Wens}
\email[]{vwens@ulb.ac.be}
\affiliation{Laboratoire de Cartographie fonctionnelle du Cerveau, UNI -- ULB Neurosciences Institute, Universit\'e libre de Bruxelles (ULB), Brussels, Belgium}


\begin{abstract}

Network theory and inverse modeling are two standard tools of applied physics, whose combination is needed when studying the dynamical organization of spatially distributed systems from indirect measurements.
However, the associated connectivity estimation may be affected by spatial leakage, an artifact of inverse modeling that limits the interpretability of network analysis.
This paper investigates general analytical aspects pertaining to this issue.
First, the existence of spatial leakage is derived from the topological structure of inverse operators.
Then, the geometry of spatial leakage is modeled and used to define a geometric correction scheme, which limits spatial leakage effects in connectivity estimation.
Finally, this new approach for network analysis is compared analytically to existing methods based on linear regressions, which are shown to yield biased coupling estimates.




\end{abstract}

\maketitle 

\pagestyle{fancy}
\fancyhead{}
\fancyhead[R]{\footnotesize \sc Physical Review E \textbf{91}, 012823 (2015)} 
\fancyhead[LE]{\footnotesize \sc Vincent Wens}
\fancyhead[LO]{\footnotesize \sc Investigating complex networks with inverse models}
\renewcommand{\headrulewidth}{0pt}


\section{Introduction}

Investigating the dynamical organization of spatially extended systems represents a challenging problem arising in many scientific disciplines from physics to biology and medicine.
Such systems can often be understood as networks of local units whose connectivity structure generates a range of complex collective behaviors that can not be achieved by isolated units, and which in turn determines to a large extent the functions of these systems \cite{Strogatz2001Nature}.
However, the network architecture underlying collective dynamics may not be directly observable and must rather be inferred or modeled from experimental data. 

In this context, network theory has emerged as an important phenomenological tool based on representations of complex systems as graphs \cite{Albert2002RMP,Newman2003siam, Boccaletti2006physrep}. 
Graph construction typically requires two modeling steps: (i) node selection, which consists in choosing the system variables playing the role of graph nodes, and (ii) connectivity estimation, which amounts to defining a coupling measure determining the links between these nodes \footnote{When connectivity estimation is based on some correlation between pairs of time series representing nodes activity, the resulting graph is sometimes called a \emph{functional network} to distinguish it from the actual anatomy of physical connections. This paper focuses on functional networks.}.
Both steps are known to affect the ensuing graph topology \cite{Butts2009Science, Bialonski2010Chaos} and this can lead to misinterpreting properties such as ``small-world-ness,'' a potentially fundamental organizational principle in various complex networks \cite{Watts1998SW}.
Node selection is especially difficult for spatially distributed systems, in particular when they are experimentally accessible through indirect or incomplete observations only. 
In a situation where a system is probed via sensors, one choice would be to directly use sensors as nodes.
However, this approach may be of limited value insofar as measurements miss or mix the functionally relevant system units, leading to graphs that strongly depend on the experimental setup \cite{Bialonski2010Chaos}.
A more natural method would be to first reconstruct the relevant units from experimental observations and then use them as nodes in a ``second-level'' network analysis. 
This reconstruction problem may not have a unique solution and thus require the introduction of an inverse model \cite{Tarantola2006Nature,*Jobert2013inv, Tarantola2004book}, which typically combines (i) a least-squares fit of an observational model of measurements 
(direct model) 
to the actual experimental data, with (ii) constraints enforcing extra \emph{a priori}, data-independent hypotheses about the system (priors) and leading to a unique solution. 
Priors often include assumptions about the spatial distribution of the system, which yields reconstructions with spatial properties characteristic of the inverse model rather than of the underlying configuration. 
This spatial deformation effect induces artifacts in the subsequent connectivity estimation and thus fundamentally limits the integration of inverse modeling and network analysis. 
We shall refer to this effect as \emph{spatial leakage} \cite{*[{There are other sources of biases in connectivity estimation unrelated to spatial leakage, but we shall not consider them here. For example, for biases induced by the finiteness of data, see }] [{}] Bialonski2011plos, *Zalesky2012corr}, which is the main subject of this paper.

\begin{figure}
\includegraphics[width=8.5cm]{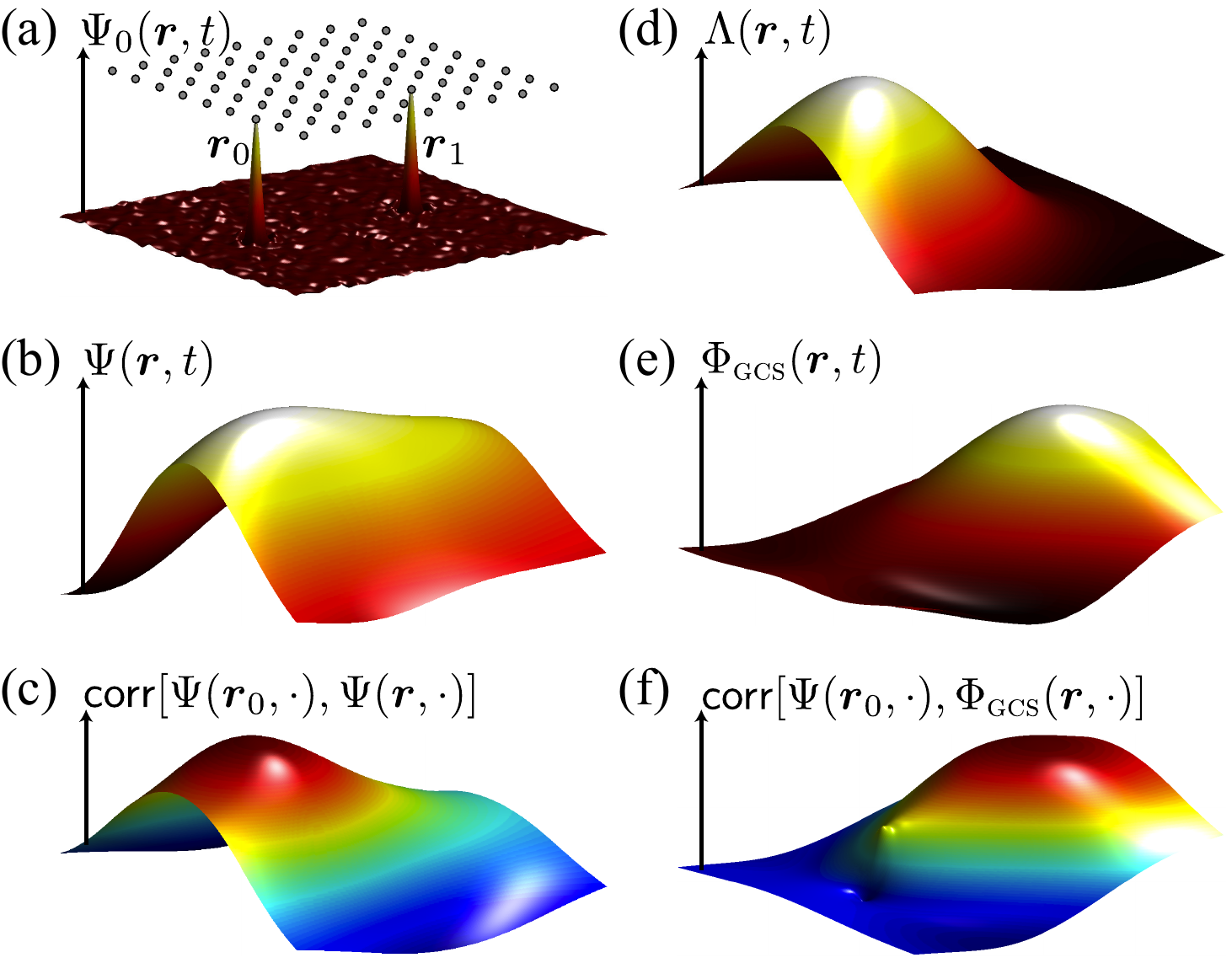}
\caption{\label{fig1} 
\emph{Spatial leakage.} (a) Snapshot of a real scalar field $\Psi_0(\br,t)$ representing two correlated sources at $\br_0$ and $\br_1$. The dots indicate sensors. (b) Its reconstruction $\Psi(\br,t)$ based on least-$\mathscr L_2$-norm estimation. (c) Temporal correlation map with node $\br_0$ as seed and directly estimated from the reconstruction. (d) Model $\Lambda(\br,t)$ of spatial leakage from $\br_0$, and (e) resulting corrected field $\Phi_\gcs(\br,t)$ used in the geometric correction scheme. (f) Temporal correlation map with seed $\br_0$ obtained after correction.}
\end{figure}

\begin{center} {\it\small \textbf{Example and motivation}} \end{center}
To illustrate the notion of spatial leakage more concretely, let us consider the simulation shown in Fig.~\ref{fig1}.  
Figure \ref{fig1}(a) depicts the surface charge density $\Psi_0(\br,t)$ of an electrically charged two-dimensional sheet, as well as a sensors array (gray dots) sampling the resulting electric field away from the sheet. 
It is assumed that the charge distribution varies slowly enough in time so Coulomb's inverse-square law applies. 
This defines the direct model of this setup, which allows prediction of electric field measurements $\boldsymbol E(\boldsymbol R_i,t)$ at sensors location $\boldsymbol R_i$ from given surface charge density $\Psi_0(\br,t)$ via
\be\label{coulomb2d} \boldsymbol E(\boldsymbol R_i,t)=\frac{1}{4\pi\varepsilon_0}\, \int_{\textrm{sheet}} \d^2\br\,  \frac{\boldsymbol R_i-\br}{\|\boldsymbol R_i-\br\|^3}\, \Psi_0(\br,t)\, .\ee
The simulated configuration consists of two pointlike charges placed at sheet locations $\br_0$ and $\br_1$ in a slightly noisy background; the two charges being temporally covarying. 
This system therefore represents a network of two charges which features in particular no local clustering \cite{Watts1998SW} (e.g., the charge density fluctuations at $\br_0$ are temporally independent of those in its neighborhood, where only background noise contributes).

We now consider the inverse question of inferring this network structure on the basis of the sensors data, assuming that they were generated via Eq.~\eqref{coulomb2d} but without further knowledge of the surface charge density.
This is an example of network analysis requiring inverse modeling before connectivity estimation.
Indeed, the problem of estimating the surface charge density from the known data is undetermined (since sensors only probe the electric field on a finite array of locations $\boldsymbol R_i$) and an inverse model thus should be used.
Without any further information, a quite natural prior would be to assume that the surface charge density is a spatially uniform (i.e., no sheet location \emph{a priori} singled out), temporally uncorrelated (i.e., no \emph{a priori} connectivity) Gaussian noise.
An inverse model embodying this prior is the least-$\mathscr L_2$-norm estimate \cite{Tarantola2004book} (see Sec.~\ref{invop_analytics}) and the resulting surface charge density reconstruction $\Psi(\br,t)$ is depicted in Fig.~\ref{fig1}(b).
It appears to be a spatially smoothed version of the true configuration; the rough reason being that the inverse model tries to enforce spatial uniformity while keeping good consistency between the known data and the electric field predicted by the reconstruction (i.e., Eq.~\eqref{coulomb2d} with $\Psi(\br,t)$ in place of $\Psi_0(\br,t)$).
This spatial deformation has an important effect on the subsequent connectivity estimation, to which we now turn. 
To investigate the network structure, a coupling measure must first be chosen (e.g., temporal correlation) and then estimated between the reconstructed charge density fluctuations of each pair of sheet locations, from which a graph can be derived.
A visually clearer approach is to use seed-based connectivity maps, i.e., spatial maps of coupling estimates between a chosen ``seed'' location to elsewhere (thus describing only parts of the full graph).
Figure \ref{fig1}(c) shows the resulting temporal correlation map from the seed $\br_0$ (which is singled out from Fig.~\ref{fig1}(b) as the global maximum). 
Its main feature is the significant spread of correlation from the trivially maximal (unit) self-correlation at $\br_0$. 
This spread is actually due to the spatial smoothness of the reconstruction and represents spatial leakage emanating from $\br_0$.
Crucially, the location $\br_1$ is not singled out in Fig.~\ref{fig1}(c), indicating that the true connectivity between $\br_0$ and $\br_1$ is hidden by the dominating spatial leakage effect.
In terms of graph topology, the reconstructed network exhibits high local clustering (contrary to the underlying network) because of this strong, widespread spurious connectivity.

This example motivates the necessity of eliminating, or at least dampening, spatial leakage effects when combining inverse modeling with network analysis.
A new correction method for seed-based connectivity mapping will be presented in this paper.
A natural step for defining a principled approach is to first understand the mechanisms underlying spatial leakage, which originates (as hinted above) from topological-like properties of inverse models such as their spatial continuity. 
In the first part of this work, we shall therefore analyze in some details the topological structure of spatial leakage.
We shall then present a new \emph{geometric correction scheme} (GCS) derived from an explicit model of spatial leakage geometry.

In the interest of illustration, let us precede our results and apply this method to the example shown in Fig.~\ref{fig1}. Figure \ref{fig1}(d) depicts the model $\Lambda(\br,t)$ of spatial leakage emanating from $\br_0$ and Fig.~\ref{fig1}(e) presents the corrected reconstruction $\Phi_\gcs(\br,t)$ in which this spatial leakage contribution has been eliminated. 
The resulting seed-based connectivity map between the reconstructed charge density $\Psi(\br_0,t)$ at $\br_0$ and the corrected charge density $\Phi_\gcs(\br,t)$ is shown in Fig.~\ref{fig1}(f) and now exhibits a correlation maximum at the location $\br_1$. 
The corrected connectivity map is clearly not perfect, since the peak at $\br_1$ appears much smoother than expected. 
This is because the method corrects for spatial leakage effects from $\br_0$ only but does not modify other spatial properties such as smoothness around $\br_1$ (which explains the remaining blur), nor does it increase spatial resolution of the original inverse model. 
Despite this limitation, important features of the simulated network are recovered (i.e., low local clustering and coupling between $\br_0$ and $\br_1$), which considerably improves the interpretation of connectivity estimation compared to Fig.~\ref{fig1}(c).

\begin{center}{\it\small \textbf{Application to brain network mapping}}\end{center}
One field of application (which actually motivated the present work) where the type of analysis exemplified above arises is systems neuroscience and more specifically the mapping of brain networks from noninvasive electrophysiological measurements of cerebral activity.
The brain is a large network of neurons linked together via axonal connections, whose complex circuitry is to a large extent responsible for the emergence of various functional behaviors from basic stimulus detection to higher-level cognitive processes \cite{Buzsaki2006book}. 
One interesting principle of cerebral organization explored by neuroscientists is that the brain is structured into functional (i.e., dynamically generated) assemblies of neurons effectively bound together via, e.g., oscillatory collective neural firing patterns \cite{Buzsaki2004Science}. 
When this effect takes place on a macroscopic scale (compared to neurons scale), these assemblies form so-called brain functional networks, whose potentially fundamental importance for physiological and pathological cerebral activity has been suggested by experimental and simulation studies
\cite{*[{For reviews, see }] [{}] Deco2011balance, *Deco2011emerge}.
In this context, a developing theme of research is the investigation of spatiotemporal properties of brain functional networks 
via noninvasive electrophysiological measurements such as electro- and magnetoencephalography \cite{Hamalainen1993MEG}.
These experimental modalities use sensors placed on subjects scalp to sample the electric potential or magnetic field generated by local postsynaptic current flows at a neural population level \cite{Hamalainen1993MEG, Buzsaki2012NatRevNeurosci}.
Brain functional networks have been investigated both using sensors \cite{Stam2007biomedphys, Bullmore2009NatRevNeurosci} (notwithstanding the limitations of this approach \cite{Bialonski2010Chaos}) and reconstructed brain sources \cite{Schoffelen2009HBM, dePasquale2010pnas, *dePasquale2012neuron, Brookes2011pnas, Brookes2012neuroimage, Hipp2012NatNeurosci, Wens2014ClinNeurophys, Wens2014interintra}.
In the latter case, the analysis methods combined inverse modeling and connectivity estimation in a way similar to the simulation example described above (see Fig.~\ref{fig1}), with the flat sheet replaced by the circumvolved cortical sheet and electric charge density replaced by electric current density.
In particular, spatial leakage effects dominate over the physiological interactions and may hide brain functional networks (albeit somewhat less drastically than in Fig.~\ref{fig1}(c)), which is an important methodological limitation in this field.

To overcome this difficulty, researchers introduced various coupling measures insensitive to spatial leakage such as imaginary coherence \cite{Nolte2004imcoh}, phase lag index \cite{Stam2007pli,Hillebrand2012pli}, or orthogonalized amplitude correlation \cite{Brookes2012neuroimage,Hipp2012NatNeurosci}.
However, these methods are limited, as they are based on some form of linear regression and thus rely on quite generic assumptions about spatial leakage effects (e.g., their linearity).
This can lead to biased pictures of network connectivity, as can be illustrated by considering once again the simulation shown in Fig.~\ref{fig1}. 
Linear regression consists in eliminating the linear correlation structure between the reconstructed activity $\Psi(\br,t)$ and the seed activity $\Psi(\br_0,t)$ (illustrated in Fig.~\ref{fig1}(c)) before connectivity estimation. 
By construction, the resulting ``corrected'' correlation map must therefore be flat and cannot recover the simulated network, contrary to the geometric correction scheme (which derives from a direct description of spatial leakage effects rather than the elimination of \emph{all} linear correlations).
In the last part of this paper, we shall analyze more generally the biases introduced by approaches based on linear regression compared to the geometric correction scheme.

\begin{center}{\it\small \textbf{Aim and organization of the paper}}\end{center}
The purpose of this paper is to investigate the problem of spatial leakage by means of analytical methods, rather than focusing on specific simulations or applications \cite{*[{For a proof of concept on simulated and experimental neuromagnetic data, see }] [] Wens2014gcs}.
The idea is that the theory is quite general and could be applied to various scientific problems where the network structure of complex systems must be inferred from inverse modeling. 
We shall therefore consider some aspects that can be subjected to analytical reasoning.

First, we shall relate the existence of spatial leakage to topological features of inverse models, focusing for definiteness on least-$\mathscr L_p$-norm estimation (Sec.~\ref{invop_analytics}). 
Then we shall characterize the geometry of spatial leakage by introducing an explicit model and define the associated geometric correction scheme (Sec.~\ref{gcs}).
Finally, we shall compare the geometric correction scheme to the orthogonalization approach based on a linear regression \cite{Hipp2012NatNeurosci} and examine the biases that the latter induces in connectivity estimation (Sec.~\ref{ortho}). 
Taken together, these three analyses lay the foundations of a systematic theory of spatial leakage; yet each is somewhat independent of the others.
The key findings shall be summarized and discussed in the closing section (Sec.~\ref{discuss}).
Five appendices gather the technical details.

\section{Topological structure of spatial leakage}\label{invop_analytics}

\subsection{Setup and general considerations}\label{setup}

The formal setup in which the various analyses of this paper will take place is the following.

\begin{center}{\it \small \textbf{Direct modeling}}\end{center}
Let us consider a continuous dynamical system living in a domain $\mathcal D$ of a $d$-dimensional Euclidean space.
Its state is described by a field $\Psi_0(\br,t)$, which will be assumed real and scalar for notational convenience ($\br\in\mathcal D$ and $t$ denoting time).
The system is accessible to the experimenter via $m$ simultaneous measurement time series $\bm(t)$, and the physics of the problem determines the direct model, which formally consists in the knowledge of a mapping $\mathsf L$ (direct operator) predicting the time series $\bm(t)$ from a given system state $\Psi_0(\br,t)$ via
\be\label{directmodel} \bm(t)=(\mathsf L\Psi_0)(t)\, .\ee
For simplicity, the contribution of measurement noise will be neglected altogether.

In this section, we shall limit ourselves to the case where $\mathsf L$ is a linear, time-independent functional of the form
\be\label{directop} (\mathsf L\Psi_0)(t)=\int_{\mathcal D} \d^d\br\, \bL(\br)\, \Psi_0(\br,t)\, , \ee
where $\bL(\br)\in\mathbb R^m$ is the impulse-response function, i.e., $\bL(\br_0)$ represents the data obtained when the system state is concentrated at the point $\br_0$, $\Psi_0(\br,t)=\delta(\br-\br_0)$. 
We shall also assume $\bL(\br)$ to be a smooth function of $\br$ on $\mathcal D$.
For example, the direct model \eqref{coulomb2d} used in the simulation shown in Fig.~\ref{fig1} is of this form; $\mathcal D$ denoting the two-dimensional charged sheet and $\bL(\br)$ gathering the components of the electric field $\boldsymbol E(\boldsymbol R_i,t)$ (at all sensors locations $\boldsymbol R_i$) generated by a single unit charge placed at $\br$.  
Since sensors are located outside the sheet, $\br\neq\boldsymbol R_i$ and $\bL(\br)$ depends smoothly on the charge position.

\begin{center}{\it\small \textbf{Inverse modeling}}\end{center}
In an experimental situation, the observer has access to the data $\bm(t)$ but does not know the underlying system state $\Psi_0(\br,t)$.
The direct operator being typically noninvertible (e.g., the functional \eqref{directop} has an infinite-dimensional kernel, meaning that infinitely many field configurations determine the same observations), the system state (and its connectivity structure) must rather be inferred by inverse modeling \cite{Tarantola2004book}.
Very generally, an inverse model consists in the definition of a family of mappings $\mathsf W_{\br}$ (inverse operator) determining an estimate $\Psi(\br,t)$ of the system state from experimental time series $\bm(t)$ via
\be\label{invop}\Psi(\br,t)=(\mathsf W_{\br}\bm)(t)\, .\ee
An inverse operator thus plays the role of a ``regularized inverse'' of $\mathsf L$.
Many inversion schemes can be defined for a given direct model, depending on the priors used. 

\begin{center}{\it\small \textbf{Spatial leakage}}\end{center}
By definition, spatial leakage refers to biases in connectivity estimation due to spatial properties intrinsic to inverse operators, which reflect the direct model and priors rather than dynamical information about the system state \cite{*[{Spatial leakage differs from the notion of \emph{spectral} leakage, which originates from finite-dimensional truncations of the space of fields $\Psi(\br,t)$; see }] [{}] Trampert1996science}. 
By  a ``spatial property'' of a function $\Psi(\br)$ is meant an attribute such as its vanishing outside a given domain, or its continuity in variable $\br$.
An issue in analyzing spatial leakage effects is that, in general, inverse operators can not be computed explicitly, as will be exemplified below. 
Our strategy to bypass this difficulty will be based on the equivalence between (i) spatial properties of inverse operators and (ii) \emph{state-independent} spatial properties of reconstructions $\Psi(\br,t)$.
Indeed, Eq.~\eqref{invop} shows that attributes in the $\br$ dependence of $\mathsf W_{\br}$ induce (and are induced by) spatial properties of $\Psi(\br,t)$ independent of data $\bm(t)$ and thus also independent of the state $\Psi_0(\br,t)$ (since the state only appears through the direct model \eqref{directmodel}).
For example, coming back to Fig.~\ref{fig1}(b), the presence of a maximum at $\br_0$ in the reconstruction is an attribute induced by the state configuration shown in Fig.~\ref{fig1}(a), whereas its spatial smoothness is not.
In this section, we shall investigate spatial leakage by exhibiting some state-independent spatial properties.

\subsection{Least-$\mathscr L_p$-norm estimation}\label{MLpE}

To make these very general considerations more concrete, we specialize here to specific but prototypal inverse models: the least-$\mathscr L_p$-norm estimates.
Let us review their definition \cite{Tarantola2004book, Daubechies2004Lp}.

Least-$\mathscr L_p$-norm estimation is based on the optimization of two competing constraints: least-squares data fitting and $\mathscr L_p$-norm minimization.
First, the reconstructed field $\Psi(\br,t)$ should be consistent with the observations in the sense that experimental data $\bm(t)$ and theoretical data $(\mathsf L\Psi)(t)$ predicted by the direct model should be close. 
Mathematically, this amounts to minimizing the model error (or misfit)
\be\label{fitterm} \mathsf V_{\textrm{misfit}}=\frac{1}{2}\, \int\d t\, \big\| \bm(t)-(\mathsf L\Psi)(t)\big\|^2\ee
over the field $\Psi(\br,t)$ (where $\|\cdot\|$ denotes the Euclidean norm on $\mathbb R^m$).
As said above, this minimization has infinitely many solutions since $\mathsf L$ is not invertible.
It can be regularized by adding a penalty term 
\be\label{Lpterm}\mathsf V_{\textrm{prior}}=\frac{\lambda}{p}\int\d t\,\d^d\br\, |\Psi(\br,t)|^p\ee
enforcing the $\mathscr L_p$-norm minimization constraint for configurations $\Psi(\br,t)$, with $p>1$ and $\lambda>0$ being parameters to be chosen.
In terms of priors, this embodies the assumption that the system state represents a spatially homogeneous white noise distributed according to a $p$-generalized Gaussian \cite{Tarantola2004book}.
The Gaussian case $p=2$ corresponds to the priors used to obtain Fig.~\ref{fig1}(b).

To summarize, least-$\mathscr L_p$-norm estimation amounts to minimizing the sum $\mathsf V = \mathsf V_{\textrm{misfit}}+\mathsf V_{\textrm{prior}}$.
%
%
An important property is that $\mathsf V$ has a unique minimizer $\Psi(\br,t)$ \cite{Tarantola2004book, Daubechies2004Lp}, which defines the reconstruction and also determines, once expressed as a functional of data, the inverse operator $\mathsf W_{\br}$ (see Eq.~\eqref{invop}).
This minimizer must extremize $\mathsf V$ and thus satisfies the condition
%
\be\label{vareq} \frac{\delta \mathsf V_{\textrm{misfit}}}{\delta\Psi(\br,t)}+ \frac{\delta \mathsf V_{\textrm{prior}}}{\delta\Psi(\br,t)} = 0\, .\ee
In general (for $p\neq 2$), the solution of this equation cannot be derived explicitly but rather is estimated numerically using iterative algorithms \cite{Tarantola2004book, Daubechies2004Lp}. 
However, this approach does not allow systematic investigation of state-independent spatial properties.
We now introduce an analytical method based on a \emph{topological} analysis.

\subsection{Topological analysis}\label{theorem1}

This section presents and discusses the main technical results that will be used next to derive the existence of spatial leakage effects in least-$\mathscr L_p$-norm estimates.

A basic implication of the spatial uniformity prior is that the only possible spatial inhomogeneities in reconstructions must be induced by the direct model (\ref{directmodel}, \ref{directop}) fitting. 
In other words, spatial properties are strongly controlled by the impulse-response function $\bL(\br)$.
Of course, the data $\bm(t)$ modulate the precise way $\bL(\br)$ affects the $\br$ dependence of the reconstruction $\Psi(\br,t)$.
This means that, in general, its geometric features are state dependent and reflect to some extent true properties of the system state (e.g., the presence of a maximum at $\br_0$ in Fig.~\ref{fig1}(b)).
Our goal is rather to find state-independent features. 
This amounts to focus on spatial properties of $\bL(\br)$ that are sufficiently generic to hold true whatever the observed data. 
An example of such generic properties are those that are topological, i.e., that are preserved under any homeomorphism (continuous transformation with continuous inverse).
It we assume that reconstructions depend continuously on data $\bm(t)$ (i.e., that $\mathsf W_{\br}$ in Eq.~\eqref{invop} is a continuous operator, for each $\br$), then varying $\bm(t)$ does not affect topological properties, which thus determine a class of state-independent spatial properties.
The method introduced here represents a generalization of this observation.

\begin{center}{\it\small \textbf{Topological conjugacy}}\end{center}
Instead of directly tackling the analysis of spatial properties for least-$\mathscr L_p$-norm estimates $\Psi(\br,t)$, our strategy will rather be to focus on fields of the form
\be\label{smallpsi} \psi(\br,t)=\bL(\br)^\trans\boldsymbol v(t)=\sum_{k=1}^m \ell_k(\br)\, v_k(t)\, .\ee
These fields are linear superpositions of $m$ basic modes $\ell_k(\br)\, v_k(t)$, whose spatial profile is given by the $k^{\textrm{th}}$ component $\ell_k(\br)$ of $\bL(\br)$, whereas its contribution to the full combination $\psi(\br,t)$ is temporally modulated by some coefficient $v_k(t)$.
Crucially, the considered coefficients  $\boldsymbol v(t)=(v_1(t),\ldots,v_m(t))\in\mathbb R^m$ do not depend on position $\br$, but otherwise they will not be known explicitly in our subsequent analyses.
Despite this, investigating the spatial properties of $\psi(\br,t)$ 
in terms of the given impulse-response function $\bL(\br)$ is a considerably simpler task.

\begin{figure}
\includegraphics[width=8.5cm]{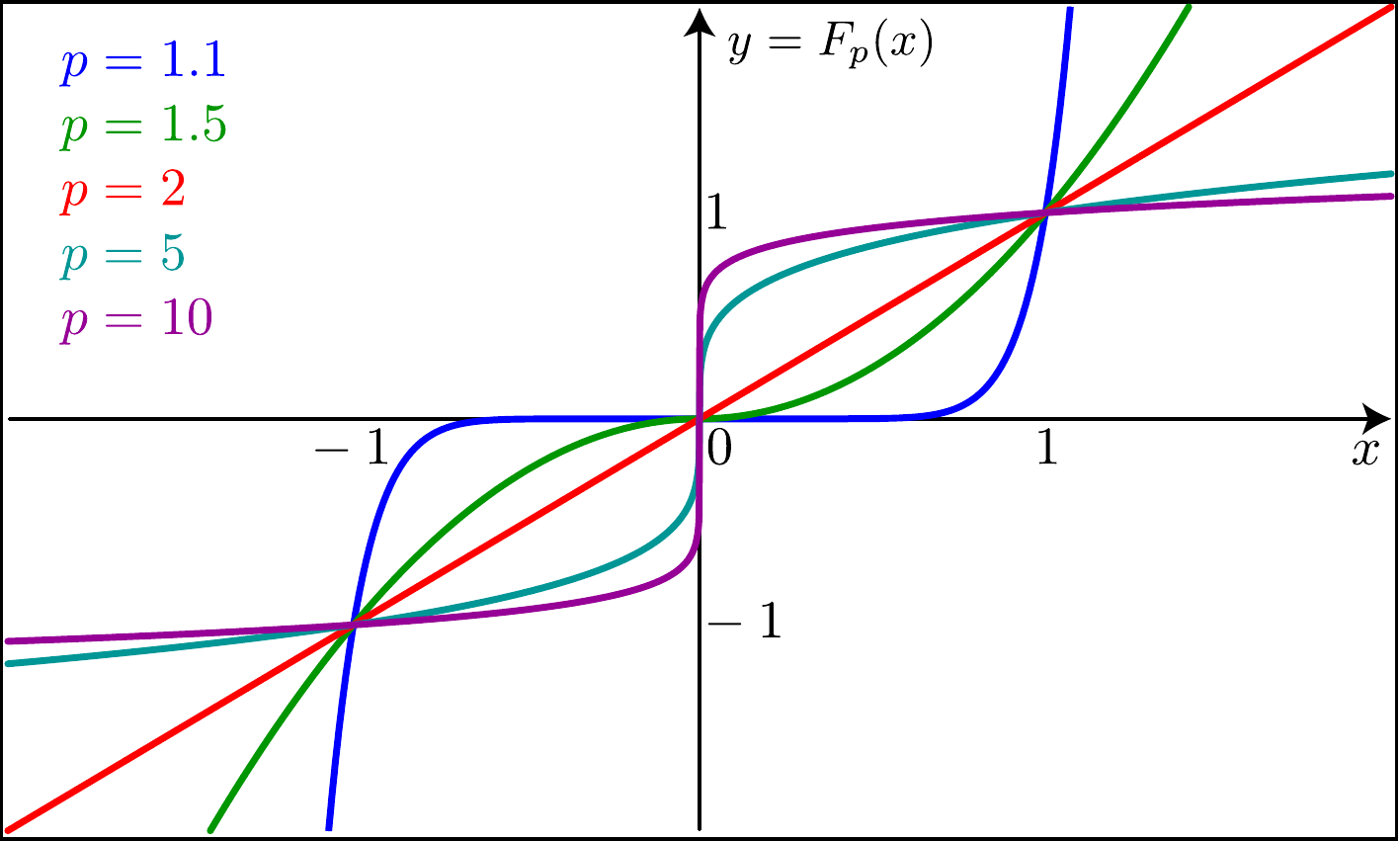}
\caption{\label{fig2} 
\emph{Topological conjugacy function $F_p$.} The graph of the homeomorphism $F_{p}$ is depicted for various values of the parameter $p$.}
\end{figure}

The justification of this strategy is based on the existence of a topological conjugacy between least-$\mathscr L_p$-norm estimates $\Psi(\br,t)$ and linear combinations $\psi(\br,t)$: For a suitable set of coefficients $\boldsymbol v(t)$ (depending on the system state), their graph can be obtained from one another via a continuous transformation.
Mathematically, this means that \emph{there exists $\boldsymbol v(t)\in\mathbb R^m$ and a homeomorphism $F_p:\mathbb R\rightarrow\mathbb R$ such that}
\be\label{conjugacy} \Psi(\br,t)=F_p(\psi(\br,t))\, .\ee
Importantly, the required homeomorphism is state independent and actually only depends on the parameter $p$. Explicitly, we have (see Fig.~\ref{fig2})
\be\label{Fp} F_p(x)=\mathsf{sign}[x]\, |x|^{1/(p-1)}\, , \ee
where $\mathsf{sign}[x]$ denotes the sign of $x$.

The above statement is the core technical result of this section and derives from the variational equation \eqref{vareq}, as proven in Appendix \ref{topoLp}. 
It implies that properties of the reconstruction $\Psi(\br,t)$ can indeed be studied via a linear combination $\psi(\br,t)$; the price to pay being to work modulo the homeomorphism \eqref{Fp}, which is the reason why the relationship is topological rather than geometric.
In principle, this suggests the following strategy to systematically explore the geometry of reconstructions: {Start with a spatial property of $\bL(\br)$, deduce what it implies for the linear combination \eqref{smallpsi} that appears in the topological conjugacy \eqref{conjugacy}, and use the transformation (\ref{conjugacy}, \ref{Fp}) to obtain the associated spatial property for $\Psi(\br,t)$.} 
An issue for the practical applicability of this method is that the corresponding coefficients $\boldsymbol v(t)$ cannot be computed explicitly in general (Appendix \ref{topoLp}).
Furthermore, these coefficients must depend on data $\bm(t)$ and thus the second step makes the resulting spatial property for $\Psi(\br,t)$ state dependent. 
To bypass these two limitations, we now focus on features of $ \bL(\br)$ that are generic enough to hold for \emph{all} linear combinations $\bL(\br)^\trans\boldsymbol v$, i.e., for any coefficients $\boldsymbol v$.
This results into the following principle, which is proven in Appendix \ref{topoLp}: {\it A spatial property valid for linear combinations $\bL(\br)^\trans\boldsymbol v$ with arbitrary $\boldsymbol v\in\mathbb R^m$ determines (via $F_p$) a state-independent spatial property for  $\Psi(\br,t)$.}


\begin{center}{\it\small \textbf{Two state-independent spatial properties}}\end{center}
We shall restrict ourselves to the demonstration of examples relevant for spatial leakage:

\begin{enumerate}[(i)]
\itemsep -0.3em
\item {\it Least-$\mathscr L_p$-norm estimates $\Psi(\br,t)$ are spatially continuous wherever $\bL(\br)$ is continuous.}
\item {\it $\Psi(\br_0,t)$ and $\Psi(\br_1,t)$ are temporally correlated whenever the vectors $\bL(\br_0)$ and $\bL(\br_1)$ are collinear.}
%
\end{enumerate}
%

%
%
%

\noindent These statements are direct applications of the above principle. 
For (i), we note that continuity is a spatial property valid for any linear combination $\bL(\br)^\trans \boldsymbol v$ (since each of the $m$ components of $\bL(\br)$ are assumed continuous) and thus induces a state-independent property for $\Psi(\br,t)$ by the above principle.
This induced property is simply the continuity of $\Psi(\br,t)$ in $\br$, as it follows from the continuity of $\psi(\br,t)=\bL(\br)^\trans \boldsymbol v(t)$ in $\br$. 
(Indeed, $\Psi(\br,t)$ is the composition $(F_p\circ \psi)(\br,t)$ of $\psi(\br,t)$ with the continuous map $F_p(x)$).
For (ii), we start from the collinearity condition $\bL(\br_1)=k\, \bL(\br_0)$ for some $k\in\mathbb R$.
This implies $\bL(\br_1)^\trans\boldsymbol v=k\, \bL(\br_0)^\trans\boldsymbol v$ for arbitrary $\boldsymbol v$, and thus produces a state-independent property for $\Psi(\br,t)$, which is determined by plugging  
\begin{equation*} \psi(\br_1,t)=\bL(\br_1)^\trans\boldsymbol v(t) = k\, \bL(\br_0)^\trans\boldsymbol v(t) = k\, \psi(\br_0,t) \end{equation*}
into the relation \eqref{conjugacy} and using the identity
\be\label{Fhomog} 
F_p(kx)=F_p(k)\, F_p(x)\, . \ee
This yields $\Psi(\br_1,t)=F_p(k)\, \Psi(\br_0,t)$, hence proving their temporal correlation.

\subsection{Existence of spatial leakage effects}\label{spatialleakage}

The above state-independent spatial properties establish the \emph{qualitative existence} of spatial leakage effects in least-$\mathscr L_p$-norm estimation, as we now discuss.

\begin{center}{\it\small \textbf{Continuity}}\end{center}
We motivated in Sec.~\ref{setup} the assumption that $\bL(\br)$ is a continuous function (see the discussion below Eq.~\eqref{directop}). 
Example (i) of Sec.~\ref{theorem1} established the fundamental analytical fact that continuity always holds for least-$\mathscr L_p$-norm estimates.
As was hinted at in the Introduction (when comparing Figs.~\ref{fig1}(b) and \ref{fig1}(c)), this state-independent property explains the existence of the strong spatial leakage effect exemplified in Fig.~\ref{fig1}(c).
Indeed, continuity at $\br_0$ forces the reconstructed time series $\Psi(\br_0,t)$ and $\Psi(\br,t)$ at two sufficiently close locations $\br_0$ and $\br$ to be almost identical, $\Psi(\br,t)\approx\Psi(\br_0,t)$.
Consequently, connectivity estimation between $\Psi(\br_0,t)$ and $\Psi(\br,t)$ is bound to yield (almost) maximal coupling in a neighborhood of $\br_0$.
This holds whatever the coupling measure, model parameters $p$, $\lambda$, and, crucially, whatever the data and system state. 
Furthermore, continuity in $\br$ implies that this spuriously high connectivity smoothly decreases as $\br$ gets away from $\br_0$, which explains the spreading effect observed in Fig.~\ref{fig1}(c).
In graph-theoretic terms, this induces falsely high clustering at node $\br_0$. 

This spatial leakage effect emanating from $\br_0$ can be characterized as local, since it happens in a connected region surrounding the seed $\br_0$. 
However, it is noteworthy that its \emph{geometric} aspects such as the shape and size of this region (which typically depend on location $\br_0$ and parameters $p$, $\lambda$) have not been estimated.
This is a limitation of the topological analysis, which works up to the homeomorphism \eqref{Fp}.
In Sec.~\ref{gcs}, we shall remedy to this situation and derive an explicit model for the geometry of this spatial leakage effect.

\begin{center}{\it\small \textbf{Spatial correlations}}\end{center}
In seed-based connectivity maps, the above spread of spurious coupling from the seed $\br_0$ decreases as $\br$ moves away from $\br_0$, but the coupling estimate may rise again once far enough.
This rebound in connectivity may indicate a true long-distance interaction but may also reflect a spatial leakage effect. 
We now exhibit a spatial property generating such long-range spatial leakage on the basis of the correlation structure of $\bL(\br)$ characterized by the normalized scalar products 
\begin{equation*}\label{Cdef} C(\br_1,\br_0)=\frac{\bL(\br_1)^\trans\bL(\br_0)}{\|\bL(\br_1)\|\, \|\bL(\br_0)\|}\, \cvd \end{equation*}

The essence of this effect can be captured by considering the idealized possibility of perfect correlation $|C(\br_1,\br_0)|=1$ for well-separated locations $\br_1$ and $\br_0$.
This condition implies the collinearity of the impulse-response vectors $\bL(\br_0)$ and $\bL(\br_1)$, i.e., $\bL(\br_1)=k\, \bL(\br_0)$ for some multiplicative factor $k$.
Example (ii) of Sec.~\ref{theorem1} showed that this ensures perfect temporal covariation of $\Psi(\br_0,t)$ and $\Psi(\br_1,t)$ (whatever the model parameters and data), which represents the sought long-range spatial leakage.
This effect can be understood more intuitively by considering the case $k=1$. Physically, the condition $\bL(\br_0)=\bL(\br_1)$ means that the two pointlike configurations $\Psi_0(\br,t)=\delta(\br-\br_0)$ and $\Psi_0(\br,t)=\delta(\br-\br_1)$ generate exactly the same measurement patterns via direct modeling (\ref{directmodel}, \ref{directop}). 
In this scenario, the direct model cannot discriminate the locations $\br_0$ and $\br_1$ (as measurements are the same) and neither does the prior (based on spatial uniformity).
In other words, the inverse model is symmetric under exchanges $\br_0\longleftrightarrow\br_1$ and must therefore lead to identical reconstructions, $\Psi(\br_1,t)=\Psi(\br_0,t)$, and spuriously maximal connectivity between $\br_0$ and $\br_1$.

It should be emphasized that the above idealization scarcely happens in practice and long-range spatial leakage effects are typically milder than the local ones (because of lower correlation values $|C(\br_1,\br_0)|$ at long distances).
Still, long-range spatial leakage has been observed \cite{Brookes2012neuroimage} and the above analysis presents a basic mechanism for its existence.
In graph-theoretic terms, this effect can lead to an underestimation of the path length, which---combined with the overestimation of clustering by local spatial leakage---might imply a spuriously small-world property of the network \cite{Watts1998SW}.

%

\subsection{Beyond least-$\mathscr L_p$-norm estimation}\label{beyondLp}

The topological analysis introduced in this section was specifically derived in the context of least-$\mathscr L_p$-norm estimation with $p>1$, but it is noteworthy that these ideas may generalize to other classes of inverse models (presumably with functions $F$ different from $F_p$).
To back up this claim, we note for example that a conjugacy also holds (with the trivial homeomorphism $F(x)=x$) for any inverse model based on \emph{unitary invariant} constraints, as described in some detail in Appendix \ref{unitaryinv}.
More generally, it would be interesting in future research to extend this analysis to other existing inverse models such as, e.g., least-mixed-norm estimates \cite{Malioutov2005mixed, Ou2009mixed, *Gramfort2013mixed}, spatial filtering \cite{Vanveen1988beamform} or Bayesian inference \cite{Wipf2009bayes, *Wipf2010bayes}.

\section{Geometry of spatial leakage}\label{gcs}

\subsection{Aim and preliminary definition}\label{prelim}

We argued in the previous section that the qualitative existence of spatial leakage can be associated with some state-independent spatial properties of reconstructions 
but we did not characterize quantitative aspects, which are more difficult to assess explicitly as they depend on the details of the inverse model. 
In practice, however, geometric information about spatial leakage effects (such as their extent) is at least as relevant as the knowledge of their qualitative existence. 

We now turn to such a geometric characterization using an analytical model of spatial leakage effects.
Specifically, we shall focus on those effects emanating from a given seed $\br_0$ and describe how true state dynamics there affects the reconstructed activity elsewhere (hence leading to spurious coupling).
An important by-product will be to introduce a new spatial leakage correction method for connectivity estimation. 
The building block of our analysis rests on the so-called resolving operator, whose definition we now describe in the very general setup of Sec.~\ref{setup} (Eqs.~\eqref{directmodel} and \eqref{invop}).

\begin{center}{\it\small \textbf{Resolving operator}}\end{center}
A basic method to characterize the geometry of an inverse operator $\mathsf W_{\br}$ (and its spatial leakage) consists in considering the reconstruction of state configurations concentrated at some location $\br_0$,
\be\label{pointlikepsi} \Psi_0(\br,t)=\delta(\br-\br_0)\, f(t)\, ,\ee
the function $f(t)$ describing their temporal activity.
Inverse modeling generally yields a deformed representation of this pointlike pattern, reflecting the structure of the inverse operator only.
(An example is given by Fig.~\ref{fig1}(d), which is the least-$\mathscr L_2$-norm estimate of a configuration with a single peak at $\br_0$, i.e., Fig.~\ref{fig1}(a) without the second peak at $\br_1$.)
Exploring reconstructions of \eqref{pointlikepsi} with varying seed $\br_0$ and activity $f(t)$ thus allows quantitative assessment of the inverse model geometry.
Most applications of this idea disregard dynamics and focus on $f(t)=1$ (which defines the point-spread function of $\br_0$ \cite{*[{Point-spread functions are often used to assess the localization biases and spatial resolution of inverse models. For example, see }] [{}] Sekihara2005locbias}) but we shall consider the general case.

Mathematically, the reconstruction $\Psi(\br,t)$ of the state \eqref{pointlikepsi} is obtained by first computing the data predicted by direct modeling \eqref{directmodel} and then applying the inverse operator \eqref{invop}, which leads to $\Psi(\br,t) = (\mathsf W_{\br}\mathsf L\Psi_0)(t)$.
As said above, it is useful to think of it as being parameterized by $\br_0$ and $f(t)$, letting them be arbitrary.
In particular, as a functional of $f(t)$, it defines the Backus-Gilbert resolving operator $\mathsf R_{\br,\br_0}$ \cite{BackusGilbert1970resolving}.
Explicitly,
\be\label{resop} (\mathsf R_{\br,\br_0}f)(t) = (\mathsf W_{\br}\mathsf L\Psi_0)(t)\, ,\ee
where $\Psi_0(\br,t)$ is given by Eq.~\eqref{pointlikepsi}. 
This operator has been used in previous publications to assess spatial leakage numerically \cite{Liu2002localization} but will serve here as the basic tool to build an analytical model. 

\subsection{Model of spatial leakage}\label{model}

The conceptual importance of the definition \eqref{resop} lies in the fact that the resolving operator $\mathsf R_{\br,\br_0}$ 
encapsulates the geometry of spatial leakage effects from the seed $\br_0$.
Indeed, $\mathsf R_{\br,\br_0}$ describes the way state activity localized at $\br_0$ contributes to the reconstruction at $\br$. 
This transfer being purely structural (the resolving operator is state independent), it induces spurious temporal codependencies between reconstructed time series at $\br_0$ and $\br$ (spatial leakage effect).
Equation \eqref{resop} is therefore a natural starting point to geometrically characterize spatial leakage. 
The main modeling step is to suitably choose the (\emph{a priori} unknown) function $f(t)$ so $(\mathsf R_{\br,\br_0}f)(t)$ reproduces spatial leakage from $\br_0$ to $\br$. 

\begin{center}{\it\small \textbf{Pointlike state}}\end{center}
To derive an estimate for $f(t)$, let us first come back to the case of pointlike state configurations \eqref{pointlikepsi}. 
We consider the location $\br_0$ known but want to infer $f(t)$ on the basis of the reconstruction, which can be written $\Psi(\br,t)=(\mathsf R_{\br,\br_0}f)(t)$ by definition of the resolving operator \eqref{resop}.
Now, it happens that $\mathsf R_{\br,\br_0}$ is often invertible when $\br=\br_0$ and we shall assume so. 
Then $f(t)$ can be obtained by inverting the relation $\Psi(\br_0,t)=(\mathsf R_{\br_0,\br_0}f)(t)$, and this leads to the \emph{exact} expression
\be\label{infer-f} f(t)=\big(\mathsf R_{\br_0,\br_0}^{-1}\Psi(\br_0,\cdot)\big)(t)\, . \ee
The reason why state activity at $\br_0$ could be inferred is that the reconstructed time series there was not mixed up with other dynamical processes taking place elsewhere and thus only appeared as a version of $f(t)$ filtered by direct and inverse modeling. 
This function was obtained by ``defiltering'' $\Psi(\br_0,t)$ (using $\mathsf R_{\br_0,\br_0}^{-1}$).

\begin{center}{\it\small \textbf{The model}}\end{center}
The above reasoning leading to Eq.~\eqref{infer-f} has no reason to hold in general since no assumption of the form \eqref{pointlikepsi} can usually be made about the unknown state configuration $\Psi_0(\br,t)$.
Nevertheless, we propose to use Eq.~\eqref{infer-f} as a rough model for the required function $f(t)$.
We shall therefore describe the geometric distribution of spatial leakage effects from $\br_0$ in terms of the field configuration
\be\label{Lambda} \Lambda(\br,t)=\big(\mathsf R_{\br,\br_0}\mathsf R_{\br_0,\br_0}^{-1}\Psi(\br_0,\cdot)\big)(t)\, ,\ee
which is nothing but $(\mathsf R_{\br,\br_0}f)(t)$ together with Eq.~\eqref{infer-f}.
More precisely, the central claim of this section is that 
{\it the reconstruction can be approximately decomposed as
\be\label{Psi-dec} \Psi(\br,t)\approx\Phi(\br,t)+\Lambda(\br,t)\, , \ee
where $\Phi(\br,t)$ represents the reconstruction free from the spatial leakage effects emanating from $\br_0$.} 

The justification of this model (beyond the case of pointlike states, for which $\Phi(\br,t)=0$) rests on two approximations: 

\begin{enumerate}[(i)]
\itemsep -0.3em
\item {\it Spatial leakage effects to $\br_0$ are neglected and}
\item {\it spatial leakage effects from $\br_0$ are linearized.}
\end{enumerate}

%
%
%

\noindent The first ensures that Eq.~\eqref{Lambda} correctly estimates the geometric distribution of spatial leakage emanating from $\br_0$, whereas the second explains its additive contribution to the reconstruction in Eq.~\eqref{Psi-dec}. 
We now discuss these two conditions in more detail.
A formal derivation of this model is given in Appendix \ref{model-derivation}.

\begin{center}{\it\small \textbf{Approximation (i)}}\end{center}
This condition originates from the comment made below Eq.~\eqref{infer-f} that the ``defiltering'' step rightly estimates the underlying activity $f(t)$ at $\br_0$ only if no other true dynamical process significantly contributes to the reconstruction at $\br_0$. 
Indeed, an additional process located at some $\br\neq\br_0$ will generally influence the reconstructed activity $\Psi(\br_0,t)$ via a spatial leakage effect from $\br$ to $\br_0$, which would then invalidate the relation $\Psi(\br_0,t)=(\mathsf R_{\br_0,\br_0}f)(t)$ and the ensuing estimates (\ref{infer-f}, \ref{Lambda}). 
Their approximate validity therefore requires that spatial leakage effects from other activated regions to $\br_0$ be negligible. 

For least-$\mathscr L_p$-norm estimation, this condition can be reinterpreted somewhat more intuitively as the spatial isolation of the node $\br_0$, i.e., the system state should be of the form \eqref{pointlikepsi} in some neighborhood of $\br_0$.
This is because its spatial leakage is dominated by local effects decreasing with distance (Sec.~\ref{spatialleakage}), so the reconstruction at $\br_0$ is mainly affected by state activity located not too far away.
However, this interpretation breaks down when significant long-range spatial leakage exists.
In any case, this condition is less severe than the pointlike state assumption \eqref{pointlikepsi}.

\begin{center}{\it\small \textbf{Approximation (ii)}}\end{center}
The second approximation concerns the way spatial leakage from $\br_0$ affects the reconstruction elsewhere.
Indeed, although the definition \eqref{Lambda} estimates how true activity $f(t)$ at $\br_0$ transfers to locations $\br\neq\br_0$, its contribution to $\Psi(\br,t)$ is generally nonlinear.
On the other hand, the model \eqref{Psi-dec} assumes an additive effect and thus entails a linearization of spatial leakage. 

This approximation may be expected to hold relatively well for local spatial leakage effects arising from the (state-independent) spatial continuity of reconstructions, as is the case for least-$\mathscr L_p$-norm estimates (Sec.~\ref{spatialleakage}).
Indeed, continuity implies that
\begin{equation*} \Psi(\br,t) \approx \Psi(\br_0,t) = \Lambda(\br_0,t) \approx \Lambda(\br,t)\end{equation*}
up to terms vanishing as $\br\rightarrow\br_0$. (The middle equality follows by setting $\br=\br_0$ in Eq.~\eqref{Lambda}.) 
The spatial leakage distribution \eqref{Lambda} therefore contributes linearly to the reconstruction in a neighborhood of $\br_0$ (and is actually the dominant part).
It is also noteworthy that spatial leakage effects are linear for $p=2$, in which case the linearization condition is automatically satisfied.

\subsection{Geometric correction scheme}\label{gcs-def}

The above model suggests a spatial leakage correction method for connectivity estimation between the reconstructed time series $\Psi(\br_0,t)$ and $\Psi(\br,t)$.
Indeed, according to Eq.~\eqref{Psi-dec}, spatial leakage from $\br_0$ can be eliminated from the reconstruction $\Psi(\br,t)$ by subtraction of its distribution \eqref{Lambda}.
The geometric correction scheme (GCS) thus consists in estimating connectivity between $\Psi(\br_0,t)$ and the corrected reconstruction defined by
\be\label{gcs-field} \Phi_\gcs(\br,t)=\Psi(\br,t)-\Lambda(\br,t)\, . \ee
%

The justification of this method is clearly conditional to the validity of the model and its two approximations.
In the limit where the model \eqref{Psi-dec} is exact, the field configuration $\Phi_\gcs(\br,t)=\Phi(\br,t)$ is free from spatial leakage effects emanating from $\br_0$ and the associated spurious couplings are thus absent between $\Psi(\br_0,t)$ and $\Phi_\gcs(\br,t)$. 
An important aspect is that spatial leakage effects, and only those effects, are removed from the reconstructed field. 
This implies that the ensuing connectivity estimation does not contain extra biases introduced by the method itself and only reflects true underlying connectivity (see the two properties discussed below).
It is noteworthy that this aspect is not shared by existing approaches \cite{Nolte2004imcoh, Stam2007pli, Hillebrand2012pli, Brookes2012neuroimage, Hipp2012NatNeurosci}, as will be discussed in Sec.~\ref{ortho}.
On the other hand, when the model approximations are violated, the application of the geometric correction scheme is not theoretically justified.

\begin{center}{\it\small \textbf{Two key properties}}\end{center}
For future reference, we state here two properties of the geometric correction scheme: 

\begin{enumerate}[(i)]
\itemsep -0.3em
\item {\it No correction is effectively applied at the location $\br_1$ in the absence of spatial leakage effects from $\br_0$ to $\br_1$.}
\item {\it When the spatial leakage model is valid and in the absence of interactions in the system state, no connectivity is disclosed between $\Psi(\br_0,t)$ and $\Phi_\gcs(\br,t)=\Phi(\br,t)$.}
\end{enumerate}
%

%
%
%

\noindent These two statements represent reasonable expectations for a spatial leakage correction method (but they are not satisfied in other approaches, see Sec.~\ref{ortho}).
The first means that the reconstruction is unaffected by the method wherever no correction is necessary, as it should. 
This is because the absence of spatial leakage from $\br_0$ to $\br_1$ implies the relation $\mathsf R_{\br_0,\br_1}=0$, so the model \eqref{Lambda} indeed vanishes at $\br_1$, $\Lambda(\br_1,t)=0$, and thus the corrected and uncorrected reconstructions coincide there, $\Phi_\gcs(\br_1,t)=\Psi(\br_1,t)$ by definition \eqref{gcs-field}.
The second statement means that no artifactual coupling is introduced after correction and only genuine interactions contribute to the ensuing connectivity estimation. 
This is because spatial leakage effects from $\br_0$ are precisely eliminated without further modification of the reconstruction.  
The formal derivation is relegated to Appendix \ref{model-derivation}.

\subsection{Model limitations}\label{limit}

In a given experimental situation, it may be difficult to \emph{a priori} evaluate whether the model approximations are accurate.
In any case, the geometric correction scheme could be used as a first approach to roughly eliminate (or at least dampen) spatial leakage effects in connectivity estimation.
Of course, the resulting network properties should then be considered cautiously. 
A topic for future work is to test the practical relevance of this method for specific applications \cite{Wens2014gcs}.

There are also two further limitations of the geometric correction scheme that should be noted. 
The first is that it only corrects spatial leakage effects emanating from a single seed $\br_0$.
This means that maps of connectivity estimates are still affected by other spatial leakage effects.
(For example, coming back to the simulation shown in Fig.~\ref{fig1}, the spread of connectivity around the node $\br_1$ in Fig.~\ref{fig1}(f) is due to spatial leakage effects from $\br_1$.)
The second is that we completely neglected measurement noise, which may invalidate the method in cases of poor signal-to-noise ratio. 
It would be interesting in future work to bypass these two limitations.

\section{Statistical estimation of spatial leakage}\label{ortho}

\subsection{The orthogonalization method}\label{ortho-def}

In the previous section, we introduced the geometric correction scheme, a model-based method to dampen spatial leakage effects in connectivity estimation.
Another possible approach is to derive correction methods using considerations of statistical nature.
\emph{A priori}, statistical arguments are limited because they do not rely on a structural description of spatial leakage and must thus bias connectivity estimation. 
The goal of this section is to investigate this question analytically.

Statistically based correction methods have been designed for the connectivity indices commonly applied to oscillatory systems, such as orthogonalization \cite{Brookes2012neuroimage, Hipp2012NatNeurosci} for amplitude correlation and the phase lag index \cite{Stam2007pli, Hillebrand2012pli} for phase coupling.
Among these two examples, orthogonalization is the most far-reaching because it can also be applied to phase connectivity and actually encompasses the phase lag index (see Sec.~\ref{phase}).
We shall therefore focus here on the orthogonalization method and contrast it with the geometric correction scheme. 

We begin by reviewing the instantaneous orthogonalization defined in Ref.~\cite{Hipp2012NatNeurosci}.
Technically, we shall need to consider \emph{complex-valued} fields because this method was developed for time-frequency representations (time-dependent Fourier or wavelet coefficients) to investigate nonstationary oscillatory behavior. The theory developed in Sec.~\ref{gcs} applies without change to this case.

\newpage
\begin{center}{\it\small \textbf{Linear regression model}}\end{center}
The idea of the method is to eliminate \emph{all} linear instantaneous connectivity between the reconstructed (Fourier or wavelet) time series $\Psi(\br_0,t)$ and $\Psi(\br,t)$ via a linear regression.
This amounts to decompose the reconstruction as 
\be\label{Psi-dec-orth} \Psi(\br,t)=\Lambda_{\small \bot}(\br,t)+\Phi_\bot(\br,t)\, , \ee
using notations analogous to those in Sec.~\ref{model}.
Here $\Lambda_\bot(\br,t)$ denotes the optimal model of $\Psi(\br,t)$ predicted by seed activity $\Psi(\br_0,t)$ on the basis of a linear and instantaneous (no phase lag or time delay) coupling. 
Explicitly,
\be\label{Lambda-orth} \Lambda_\bot(\br,t)=k_\bot(\br,\br_0,t)\, \Psi(\br_0,t)\, , \ee
where $k_\bot(\br,\br_0,t)$ is the real-valued regression coefficient,
\be\label{regresscoeff} k_\bot(\br,\br_0,t)=\mathsf{Re}\big[\Psi(\br,t)/\Psi(\br_0,t)\big]\, . \ee
The rest $\Phi_\bot(\br,t)$ represents the prediction error and is then obtained via Eq.~\eqref{Psi-dec-orth}.
The explicit formula \eqref{regresscoeff} follows by minimization of the amplitude squared prediction error $|\Phi_\bot(\br,t)|^2$ over $k_\bot(\br,\br_0,t)\in\mathbb R$, and equivalently enforces the relation 
\be\label{orthorel} \mathsf{Re}\big[\Phi_\bot(\br,t)/ \Psi(\br_0,t)\big] = 0\, ,\ee
see Ref.~\cite{Hipp2012NatNeurosci} for details. 
This equation shows that the phase difference between the complex numbers $\Psi(\br_0,t)$ and $\Phi_\bot(\br,t)$ is fixed to $\pm\pi/2$ at all times.
This orthogonality property embodies the fact that no linear instantaneous coupling remains between them.

\begin{center}{\it\small \textbf{Instantaneous orthogonalization}}\end{center}
The orthogonalization method interprets the regression model $\Lambda_\bot(\br,t)$ as a statistical estimate of spatial leakage effects from $\br_0$ and uses the prediction error $\Phi_\bot(\br,t)$ as corrected reconstruction.
Orthogonalized connectivity is then obtained by connectivity estimation between seed activity $\Psi(\br_0,t)$ and the orthogonalized field $\Phi_\bot(\br,t)$.

The strict validity of this approach entails some assumptions that should be emphasized: 

\begin{enumerate}[(i)]
\itemsep -0.3em
\item {\it Spatial leakage from $\br_0$ is linear,}
\item {\it spatial leakage from $\br_0$ is instantaneous, and}
\item {\it spatial leakage from $\br_0$ is the only source of linear instantaneous coupling between $\Psi(\br_0,t)$ and $\Psi(\br,t)$.}
\end{enumerate}
%

%
%
%
%

\noindent The first two reflect the hypotheses of the associated linear regression and constrain the structure of spatial leakage. 
Their validity depends on the direct and inverse models at hand (see Secs.~\ref{invop_analytics} and \ref{gcs}). 
The third is a dynamical hypothesis about the system state, which should not give rise to genuine linear instantaneous connectivity. 
This is because the linear regression can not discriminate between such true couplings and spatial leakage effects.
Orthogonalization thus eliminates them altogether and this can lead to an overcorrection bias.

\subsection{Geometric correction scheme versus orthogonalization}\label{compare}

In the remaining sections, we shall compare the geometric correction scheme (Sec.~\ref{gcs-def}) and the orthogonalization method. 
We start with a qualitative general discussion and then turn to more quantitative aspects.

\begin{center}{\it\small \textbf{Qualitative comparison}}\end{center}
Some features are common to both techniques, e.g., the linearity hypothesis (Secs.~\ref{model} and \ref{ortho-def}) or the nonelimination of spatial leakage effects other than those emanating from $\br_0$ (Sec.~\ref{limit}). 
But other hypotheses induce differences.

The main conceptual advantage of the geometric correction scheme is that it relies on a structural description of spatial leakage adapted to the inverse operator, whereas orthogonalization uses statistical features of reconstructed signals without explicit reference to the inverse model. 
This implies that the second and third hypotheses used by orthogonalization (Sec.~\ref{ortho-def}) are not necessary for the geometric correction scheme.
Indeed, the latter automatically conforms to the temporal characteristics of spatial leakage (and more precisely of the resolving operator, see Secs.~\ref{prelim} and \ref{model}), which may or may not have instantaneous effects depending on the inverse model.
Furthermore, it does not confound true connectivity with spatial leakage effects and so does not exhibit the overcorrection bias of orthogonalization.
For example, this bias implies that the reconstruction may be nontrivially affected by orthogonalization even in the absence of spatial leakage, whereas this does not happen with the geometric correction scheme (see property (i) of Sec.~\ref{gcs-def}).

On the other hand, some hypotheses of the geometric correction are not required for orthogonalization. 
First, measurement noise was neglected in our derivation of the spatial leakage model (Sec.~\ref{limit}), whereas the regression automatically takes care of noise contributions.
This represents an advantage of orthogonalization.
Second, the condition that spatial leakage effects to $\br_0$ are negligible (Sec.~\ref{model}) was not explicitly assumed in orthogonalization.
It is noteworthy, however, that this hypothesis was instrumental to characterize spatial leakage exactly and avoid any overcorrection bias in the geometric correction scheme.

\begin{center}{\it\small \textbf{Framework for quantitative analysis}}\end{center}
We now turn to some quantitative considerations, our goal being to illustrate the overcorrection effect of orthogonalization (compared to the geometric approach). 
To single out this overcorrection bias, we shall work from now on in a restricted framework where spatial leakage effects are both linear and instantaneous, so our model \eqref{Lambda} assumes the form
\be\label{Lambda-liinst} \Lambda(\br,t)=k(\br,\br_0,t)\, \Psi(\br_0,t)\, . \ee
An explicit example is given in Appendix \ref{liinst-derivations}.
The coupling constant $k(\br,\br_0,t)\in\mathbb R$ encodes the spatial leakage effect. 
It is state independent and determined by the direct and inverse models only, contrary to the regression coefficient \eqref{regresscoeff}. 
Actually, this is the \emph{sole} remaining difference between the two correction approaches.

\begin{center}{\it\small \textbf{Quantitative comparison}}\end{center}
We now compare the coupling estimates $k(\br,\br_0,t)$ and $k_\bot(\br,\br_0,t)$ that are eliminated by the geometric correction scheme and orthogonalization, respectively.
As shown in Appendix \ref{liinst-derivations}, they are related via
\be\label{krelation} k_\bot(\br,\br_0,t)=k(\br,\br_0,t)+\mathsf{Re}\big[\Phi_\gcs(\br,t)/\Psi(\br_0,t)\big]\, . \ee
This embodies the general observation that the linear regression includes both spatial leakage effects and true connectivity.
More precisely, the regression coefficient \eqref{regresscoeff} appears to be the sum of the spatial leakage model coupling and of the linear instantaneous connectivity estimate after geometric correction.
In the limit where the model is exact, this second term represents a true state interaction (see property (ii) of Sec.~\ref{gcs-def}) that is eliminated by orthogonalization (overcorrection bias).

It will be useful for our subsequent analyses to promote this relation at the level of the corrected fields.
Plugging Eq.~\eqref{krelation} into the linear regression (\ref{Psi-dec-orth}--\ref{regresscoeff}) yields
\be\label{basicrelation} \Phi_\bot(\br,t)=i\, \mathsf{Im}\big[\Phi_\gcs(\br,t)/\Psi(\br_0,t)\big]\, \Psi(\br_0,t)\, ,\ee
see Appendix \ref{liinst-derivations} for details.
Geometrically, this means that the orthogonalized field $\Phi_\bot(\br,t)$ equals to the projection of the corrected field $\Phi_\gcs(\br,t)$ onto the direction perpendicular to $\Psi(\br_0,t)$. 
A similar formula was derived in Ref.~\cite{Hipp2012NatNeurosci} with the uncorrected reconstruction $\Psi(\br,t)$ in place of the corrected one $\Phi_\gcs(\br,t)$. 
This interchangeability is indicative that the linear instantaneous spatial leakage effect has been completely eliminated by orthogonalization (notwithstanding its overcorrection effect).

\begin{center}{\it\small \textbf{Polar coordinates}}\end{center}
In the next two sections, we shall focus on two specific connectivity indices classically applied to oscillatory or narrow-band time series: phase coherence and amplitude correlation. 
Their definition and analysis will be simplified by using field coordinates emphasizing phase and amplitude fluctuations,
\be\label{Atheta}\begin{split} \Psi(\br_0,t) &= A_0(t) \, \exp i\theta_0(t)\, , \\  \Phi_\gcs(\br,t) &= A_\gcs(t) \, \exp i\theta_\gcs(t)\, , \\  \Phi_\bot(\br,t) &= A_\bot(t) \, \exp i\theta_\bot(t)\, . \end{split} \ee
In these notations, the relation \eqref{basicrelation} becomes
$$ A_\bot(t)\, e^{i \theta_\bot(t)} = i\, A_\gcs(t)\, \sin(\theta_\gcs(t)-\theta_0(t))\, e^{i\theta_0(t)}\, , $$
which directly implies 
\begin{align}\label{phaserelation} \theta_\bot(t)-\theta_0(t) &= \frac{\pi}{2}\, \mathsf{sign}[\sin(\theta_\gcs(t)-\theta_0(t))]\, , \\ \label{amplrelation} A_\bot(t) &= A_\gcs(t)\, |\sin(\theta_\gcs(t)-\theta_0(t))|\, . \end{align}
These two formulas should make the above-mentioned geometric interpretation of orthogonalization quite clear.
They form the basis of our analysis of phase and amplitude connectivity, to which we now turn.

\subsection{Phase coherence}\label{phase}
Phase coherence is a basic communication mechanism in various oscillatory networks from mechanical and electrical oscillators to neurons \cite{Buzsaki2004Science, Strogatz1994chaosbook, Pikovsky2003book}.
It measures the tendency of a pair of oscillators (characterized by their phase signal) to be phase locked, i.e., to have a phase difference $\Delta\theta(t)$ fluctuating around a definite mean value.
This information is quantified by the temporal average of $\exp i\Delta\theta(t)$. 
The spatial leakage coupling \eqref{Lambda-liinst} induces a spurious tendency towards synchronization or antisynchronization (i.e., phase lag $0$ or $\pi$) that we propose to correct using the geometric correction scheme rather than orthogonalization. 
The reason is found in the overcorrection bias that affects the orthogonalized phase coupling.
This can be analyzed using Eq.~\eqref{phaserelation}.

The phase coherence between the corrected field $\Phi_I(\br,t)$ (where $I=\gcs$ or $\bot$) and the seed reconstruction $\Psi(\br_0,t)$ will be denoted by
\be\label{plv} \zeta_I =  \big\langle \exp i(\theta_I-\theta_0) \big\rangle\, ,\ee
the brackets indicating temporal averaging.
Plugging Eq.~\eqref{phaserelation} into this definition for $I=\bot$ directly yields
\be\label{pli} \zeta_\bot = i\, \big\langle\mathsf{sign}[\sin(\theta_\gcs-\theta_0)] \big\rangle=  i\, (f_+-f_-)\, ,\ee
where $f_+$ and $f_-$ denote the fraction of time during which $\sin(\theta_\gcs(t)-\theta_0(t))$ is positive and negative, respectively.
This means that orthogonalized phase coherence $\zeta_\bot$ reduces to an asymmetry measure of the corrected phase dynamics $\theta_\gcs(t)-\theta_0(t)$ and thus only depends quite coarsely on the corresponding corrected phase coupling $\zeta_\gcs$. 
That orthogonalization has such a strong effect on phase coherence should be expected. 
Indeed, it eliminates the spurious spatial leakage (anti)synchronization effect via an orthogonal projection, which fixes the phase difference between $\Phi_\bot(\br,t)$ and $\Psi(\br_0,t)$ to $\pm\pi/2$ at all times (see Eqs.~\eqref{orthorel} and \eqref{phaserelation}) and thus discards \emph{any} (anti)synchronization (overcorrection effect). 
On the other hand, the geometric correction scheme only removes the model coupling \eqref{Lambda-liinst} and can still retain some (anti)synchronization.
This indicates that $\zeta_\gcs$ is a phase connectivity measure more sensitive than $\zeta_\bot$.

\begin{center}{\it\small \textbf{The phase lag index conjecture}}\end{center}
It is noteworthy that the right-hand side of Eq.~\eqref{pli} coincides with the so-called phase lag index \footnote{More precisely, the original definition of the phase lag index requires replacing $\theta_\gcs(t)$ by the phase $\theta(t)$ of the \emph{uncorrected} reconstruction $\Psi(\br,t)$ in the right-hand side of Eq.~\eqref{pli}.  This replacement can indeed be applied, as was noted below the basic relation \eqref{basicrelation}.}.
Interestingly, this connectivity estimate was originally introduced on quite heuristic grounds and conjectured to be insensitive to spatial leakage \cite{Stam2007pli,Hillebrand2012pli}. 
Our identification of the phase lag index with orthogonalized phase coherence $\zeta_\bot$ proves that this is correct whenever spatial leakage effects are assumed linear and instantaneous, since then orthogonalization indeed completely eliminates spatial leakage (Sec.~\ref{compare}).
On the other hand, this conjecture has no reason to hold when spatial leakage is not linear or instantaneous, since orthogonalization 
then relies on wrong hypotheses (Sec.~\ref{ortho-def}).

\subsection{Amplitude correlation}\label{amplitude}
Real-world oscillatory systems are scarcely perfectly periodic and typically display narrow-band dynamics parameterized by both phase and amplitude fluctuations.
It is thus also interesting to consider amplitude connectivity. 
Orthogonalization was originally designed for amplitude correlation \cite{Brookes2012neuroimage, Hipp2012NatNeurosci} but the effect of the overcorrection bias is actually more intricate in this case than for phase connectivity.
This question was investigated in some detail in Ref.~\cite{Hipp2012NatNeurosci} using numerical simulations of a model of coherent Gaussian signals.
We review and solve this model analytically in Appendix \ref{hippmodelderivation}.
Here we focus on a general discussion of the overcorrection bias on the basis of Eq.~\eqref{amplrelation}.

This equation shows that the orthogonalized field amplitude $A_\bot(t)$ is a nonlinear mixture of the corrected field amplitude $A_\gcs(t)$ and phase lag $\theta_\gcs(t)-\theta_0(t)$.
After orthogonalization, the amplitude correlation 
\begin{equation*}\rho_\bot=\mathsf{corr}[A_\bot,A_0]=\mathsf{corr}[A_\gcs\times |\sin(\theta_\gcs-\theta_0)| , A_0]\end{equation*} 
therefore entangles both the amplitude and phase couplings obtained after geometric correction, indicating that the estimate $\rho_\gcs=\mathsf{corr}[A_\gcs,A_0]$ has higher specificity to true amplitude connectivity than $\rho_\bot$.
Indeed, the only difference is the presence of the multiplicative factor $|\sin(\theta_\gcs-\theta_0)|$ contributing by an amplitude-phase interaction between seed amplitude $A_0(t)$ and corrected phase lag $\theta_\gcs(t)-\theta_0(t)$.
Depending on its strength, this coupling can lead to either underestimation ($\rho_\bot<\rho_\gcs$) or overestimation ($\rho_\bot>\rho_\gcs$), which illustrates the nontrivial effect of the overcorrection bias in this case.
When it is small enough, this multiplicative factor is randomly related to $A_0(t)$ and effectively acts as noise decreasing the amplitude correlation $\rho_\bot$ ($\rho_\gcs$ being unaltered). 
When it is large enough, this factor has the opposite impact and can make $\rho_\bot$ increase above $\rho_\gcs$. 

The underestimation of amplitude correlation should be intuitively expected, since by design orthogonalization eliminates more connectivity that it ought. 
The overestimation is a more surprising artifact of orthogonalization emerging when this elimination of connectivity is counterbalanced by the mixing of amplitude and phase couplings.
It can lead to the observation of significant orthogonalized amplitude correlation even in the absence of genuine amplitude correlation.
This can be demonstrated using the (drastic) example where phase dynamics is related to amplitude dynamics via
\begin{equation}\label{drasticex} \theta_\gcs(t)-\theta_0(t)=\arcsin(k A_0(t)/A_\gcs(t))\, , \end{equation}
for some coefficient $k$.
Equation \eqref{amplrelation} then implies $A_\bot(t)=|k| A_0(t)$, so we have maximal correlation $\rho_\bot=1$ after orthogonalization, independently of $A_\gcs(t)$ and $\rho_\gcs$.
Appendix \ref{hippmodelderivation} describes another (albeit less dramatic) example of this artifact.
The extent to which this arises in realistic situations remains to be evaluated \cite{Wens2014gcs}.

\section{Summary and conclusion}\label{discuss}

Spatial leakage is a general methodological limitation for the network analysis of spatially distributed dynamical systems based on imperfect inverse models. 
In this paper, we investigated this problem by means of analytical methods.
Results gave some novel insights about the structure of spatial leakage and how to correct it in connectivity estimation. 

First, we explored the state-independent nature of spatial leakage by investigating some topological features of inverse models (Sec.~\ref{invop_analytics}).
This analysis highlighted the fundamental fact that spatial leakage effects reflect the structure of the direct and inverse models and not dynamical properties of the underlying system state.
A practical implication is that existing spatial leakage correction methods based on statistical estimations (and thus dependent of the state) cannot eliminate spatial leakage effects specifically and are bound to lead to overcorrection biases in connectivity analysis.  
Second, we introduced such a structural characterization of spatial leakage by deriving an explicit analytical model of its geometry on the basis of which we defined the geometric correction scheme for connectivity analysis (Sec.~\ref{gcs}).
Contrary to the statistical approach, this method satisfies some properties expected for a reasonable spatial leakage correction (elimination of state-independent coupling only and ability of disclosing genuine system interactions).
Critically, the strict validity of the model and geometric correction is based on a couple of hypotheses, which were clearly identified but may not necessarily hold in realistic situations. 
Third and finally, we compared this method to orthogonalization, a quite general example of statistically derived correction (Sec.~\ref{ortho}).
This allowed emphasis of the inherent overcorrection biases, in particular for phase and amplitude connectivity indices. 
This analysis suggested that, even when the assumptions of our spatial leakage model are not met, the geometric correction scheme is less biased and more sensitive to true connectivity.
It can therefore be considered as an improvement over statistical methods.

These analyses open perspectives for generalizations and refinements. 
For example, the topological analysis of spatial leakage (derived in the context of a specific class of inverse models, namely least-$\mathscr L_p$-norm estimation) may become more complicated in the context of adaptive inverse models \cite{Vanveen1988beamform, Wipf2009bayes, *Wipf2010bayes}, for which spatial leakage becomes somewhat data dependent. 
In particular, the generalization to the unified Bayesian framework of Ref.~\cite{Wipf2009bayes, *Wipf2010bayes} is an interesting problem. 
Another obvious improvement would be to refine the spatial leakage model and go beyond the present limitations. 
For example, it is possible that a more detailed investigation of the various expansions used to derive the model (see Appendix \ref{model-derivation}) leads to better nonlinear models for spatial leakage.

In conclusion, this paper lays out the premise of a systematic theory of spatial leakage. 
Eventually, this theory and the associated methods should find a wide array of applications since the problem of spatial leakage arises in any investigation of the network architecture of complex systems based on indirect or incomplete measurements.


\begin{acknowledgments}
I thank X. De Ti\`ege, N. Goldman, and S. Goldman for encouraging discussions.
This work was supported by a research grant from the Belgian Fonds de la Recherche Scientifique (Convention No. 3.4811.08, FRS--F.N.R.S.) and an Action de Recherche Concert\'ee (ARC: ``\emph{Pathophysiology of brain plasticity processes in memory consolidation}'', Universit\'e libre de Bruxelles).
\end{acknowledgments}


\appendix



\section{Topological conjugacy for least-$\mathscr L_p$-norm estimates}\label{topoLp}

In this appendix, we derive the topological conjugacy (\ref{conjugacy}, \ref{Fp}) as well as the principle stated in Sec.~\ref{theorem1}.

\begin{center}{\it\small \textbf{The variational equation}}\end{center}
 We begin by extracting the topological conjugacy \eqref{conjugacy} from the extremization condition \eqref{vareq}.
First, the variation of $\mathsf V_{\textrm{misfit}}$ (Eq.~\eqref{fitterm}) with respect to $\Psi(\br,t)$ yields
\be\label{deltafitterm} \frac{\delta \mathsf V_{\textrm{misfit}}}{\delta\Psi(\br,t)} = - \bL(\br)^\trans \big(\bm(t)-(\mathsf L\Psi)(t)\big)\, , \ee
as shown by using the definition \eqref{directop} together with the identity $\delta (\mathsf L\Psi)(t) / \delta\Psi(\br,t')=\bL(\br) \delta(t-t')$. 
Likewise, the variation of $\mathsf V_{\textrm{prior}}$ (Eq.~\eqref{Lpterm}) gives
\be\label{deltaLpterm} \frac{\delta \mathsf V_{\textrm{prior}}}{\delta\Psi(\br,t)} = \lambda\,|\Psi(\br,t)|^{p-2}\,\Psi(\br,t)\, . \ee
Since the extremization condition \eqref{vareq} amounts to setting the sum of these two variations to zero, we obtain
\be\label{Lpsaddle}\lambda\,|\Psi(\br,t)|^{p-2}\,\Psi(\br,t)=\bL(\br)^\trans \big(\bm(t)-(\mathsf L\Psi)(t)\big)\, .\ee 
Now taking the absolute value yields
\begin{equation*}\label{abspsi1}\lambda\,|\Psi(\br,t)|^{p-1}=\big| \bL(\br)^\trans \big(\bm(t)-(\mathsf L\Psi)(t)\big) \big |\end{equation*}
and thus
\begin{equation*}\label{abspsi2}|\Psi(\br,t)|^{p-2}=\lambda^{-\alpha_p}\, \big| \bL(\br)^\trans \big(\bm(t)-(\mathsf L\Psi)(t)\big) \big |^{\alpha_p}\, ,\end{equation*}
where we set $\alpha_p=(p-2)/(p-1)$. 
Plugging this back into Eq.~\eqref{Lpsaddle}, we finally find
\be\label{Lpestimate1} \Psi(\br,t)=\lambda^{\alpha_p-1}\, \frac{\bL(\br)^\trans \big(\bm(t)-(\mathsf L\Psi)(t)\big)}{\big|\bL(\br)^\trans \big(\bm(t)-(\mathsf L\Psi)(t)\big)\big|^{\alpha_p}}\, \cvd \ee

This expression is equivalent to the sought Eq.~\eqref{conjugacy} if we define $\psi(\br,t)=\bL(\br)^\trans\boldsymbol v(t)$ with
\be\label{vLp} \boldsymbol v(t) = \tfrac{1}{\lambda} \big(\bm(t)-(\mathsf L\Psi)(t)\big)\ee
and $F_p:\mathbb R\rightarrow\mathbb R$ by 
\be\label{Fdef} F_{p}(x)=x/|x|^{\alpha_p}=\mathsf{sign}[x]\, |x|^{1/(p-1)}\, ,\ee
where we used the identity $1-\alpha_p=1/(p-1)$ in the second equality. 
The graph of $F_p$ shown in Fig.~\ref{fig2} indicates that it is continuous and strictly increasing. 
Actually, $F_p$ is a homeomorphism since it has a continuous inverse 
\begin{equation*} F_p^{-1}(x)=|x|^{p-2}\, x\, . \end{equation*}
We have therefore proven that Eq.~\eqref{conjugacy} establishes a topological conjugacy between $\Psi(\br,t)$ and $\psi(\br,t)$.
(This relation actually falls short of being a diffeomorphism because $F_p$ and $F_p^{-1}$ are also differentiable everywhere except at $x=0$, where $\d F_p/\d x=\infty$ for $p>2$ and $\d F_p^{-1}/\d x=\infty$ for $p<2$.)
It is noteworthy that the coefficients \eqref{vLp} are themselves expressed in terms of $\Psi(\br,t)$ and generally cannot be computed without explicit knowledge of the reconstruction.

\begin{center}{\it\small \textbf{State-independent properties}}\end{center}
We now demonstrate the principle that a spatial property valid for any linear combination $\bL(\br)^\trans\boldsymbol v$ determines a state-independent spatial property for $\Psi(\br,t)$ (Sec.~\ref{theorem1}). 

First, a spatial property valid for linear combinations $\bL(\br)^\trans\boldsymbol v$, $\forall\boldsymbol v\in\mathbb R^m$, must hold for the particular linear combination $\psi(\br,t)=\bL(\br)^\trans\boldsymbol v(t)$.
The important point is that this property for $\psi(\br,t)$ does not depend on data (and thus on the system state). 
Indeed, to any two data sets $\bm(t)$, $\bm'(t)$ correspond coefficients $\boldsymbol v(t)$, $\boldsymbol v'(t)$ and thus fields $\psi(\br,t)=\bL(\br)^\trans\boldsymbol v(t)$ and $\psi'(\br,t)=\bL(\br)^\trans\boldsymbol v'(t)$. 
Both of them must satisfy the spatial property in question, which is therefore preserved by arbitrary changes in data.
Finally, the resulting state-independent property of $\psi(\br,t)$ gets transferred to the reconstruction $\Psi(\br,t)$ via the topological conjugacy (\ref{conjugacy}, \ref{Fp}).
This yields a homeomorphic (i.e., generally deformed) state-independent spatial property for $\Psi(\br,t)$, as claimed.

\section{Unitary inverse models}\label{unitaryinv}

The topological conjugacy \eqref{conjugacy} becomes particularly simple in the case of least-$\mathscr L_2$-norm estimation, since the homeomorphism \eqref{Fp} with $p=2$ is trivial ($F_2(x)=x$).
At first sight, this simplification seems to follow from the linearity of the associated variational equation \eqref{Lpsaddle}.
However, the deeper reason is the unitary invariance of the $\mathscr L_2$ norm, as will be discussed in this appendix in greater generality. 
Unitarity is a concept of symmetry that is central in physics.
Here we define unitary inverse models and derive their geometric properties on the basis of symmetry arguments.
%
%
We will show that associated reconstructions are always equal (rather than homeomorphic) to linear combinations
\be\label{unitarypsi} \Psi(\br,t)=\bL(\br)^\trans \boldsymbol v(t)\, ,\ee
for some coefficients $\boldsymbol v(t)$ that depend on the unitary inverse model and on the system state.
This implies that the topological structure of spatial leakage for least-$\mathscr L_2$-norm estimates and general unitary models are the same.

\begin{center}{\it\small \textbf{Unitary symmetry}}\end{center}
Let us begin by explaining what is meant by a unitary invariant inverse model.
It is convenient to introduce the standard inner product 
\be\label{L2scalarprod} \langle\Psi_1|\Psi_2\rangle = \int\d^d\br\, \Psi_1(\br)\, \Psi_2(\br)\, , \ee
which allows writing the compact expression $\mathsf L\Psi = \langle\bL|\Psi\rangle$ for the direct operator \eqref{directop}. 
It is then quite natural to consider unitary transformations $\Psi\rightarrow\mathsf U\Psi$ acting on fields $\Psi(\br)$ in the fundamental representation
\be\label{Ufundrep} (\mathsf U\Psi)(\br)=\int\d^d\br'\, U(\br,\br')\, \Psi(\br') \ee 
and leaving the inner product invariant, i.e., $\langle\mathsf U\Psi_1|\mathsf U\Psi_2\rangle$ is equal to $\langle\Psi_1|\Psi_2\rangle$.
Since we consider here real-valued fields, this imposes the orthogonality constraint
\be\label{Uortho} \int\d^d\br\, U(\br,\br') \, U(\br,\br'')=\delta(\br'-\br'')\, .\ee
The presence of the kernel $\bL(\br)$ explicitly forbids unitary transformations to be symmetries of the direct and inverse problems. 
A standard trick is to promote $\bL(\br)$ to a \emph{background} field transforming in the same representation. 
This restores symmetry in the direct model since then $\langle\mathsf U\bL|\mathsf U\Psi\rangle=\langle\bL|\Psi\rangle$.
Unitary inverse models are inverse models that are invariant under simultaneous transformations $\Psi\rightarrow\mathsf U\Psi$, $\bL\rightarrow\mathsf U\bL$. 

\begin{center}{\it\small \textbf{Symmetry argument}}\end{center}
This symmetry principle strongly constraints the functional dependence of reconstructions in the impulse-response function $\bL(\br)$. 
%
%
%
Very generally, the reconstruction is a functional 
\be\label{psigeneral} \Psi = \mathsf F[\ell_1,\ldots,\ell_m]\ee 
of the $m$ components $\ell_k(\br)$ of $\bL(\br)$. 
(We discard explicit reference to other dependencies such as data $\bm(t)$ and inverse model parameters, which will not play any role since they are assumed unitary invariant.)
Unitary symmetry imposes that if $\Psi$ is the reconstruction for given functions $\ell_k$, then the reconstruction for the transformed functions $\mathsf U\ell_k$ must be $\mathsf U\Psi$.
Mathematically, this translates into the property that $\mathsf F$ is \emph{unitary equivariant}, i.e., for any transformation $\mathsf U$ (\ref{Ufundrep}, \ref{Uortho}), we have
\be\label{equivF} \mathsf F[\mathsf U\ell_1,\ldots,\mathsf U\ell_m]=\mathsf U\mathsf F[\ell_1,\ldots,\ell_m]\, . \ee
This assumption imposes that the functional $\mathsf F$ must be of the form
\be\label{Fcov} \mathsf F[\ell_1,\ldots, \ell_m]=\sum_{k=1}^m \ell_k\, v_k\big(\langle\bL|\bL^\trans\rangle\big)\, , \ee
as expected from classical invariant theory.
Indeed, functions $v_k$ of the $m\times m$ matrix $\langle\bL|\bL^\trans\rangle$ of inner products $\langle\ell_j|\ell_k\rangle$ are the most general unitary invariant functionals of the $\ell_k$s, whereas the factors $\ell_k$ in Eq.~\eqref{Fcov} are necessary for this sum to transform in the fundamental representation.
All unitary invariant model parameters, including time $t$ and data $\bm(t)$, enter Eq.~\eqref{Fcov} through the $v_k$s.
Combining Eqs.~\eqref{psigeneral} and \eqref{Fcov} yields the sought Eq.~\eqref{unitarypsi}, with the $k^{\textrm{th}}$ component of $\boldsymbol v(t)$ given by $v_k$. 
These coefficients are independent of $\br$, since $\bL(\br)$ only appears via the inner products $\langle\bL|\bL^\trans\rangle$. (Strictly speaking, we also assumed that none of the unitary invariant model parameters depend on $\br$.)

\begin{center}{\it\small \textbf{Extremizing invariant functionals}}\end{center}
Another derivation that is closer to the analysis performed in Appendix \ref{topoLp} is the following.
As for least-$\mathscr L_p$-norm estimation, we suppose that the reconstruction $\Psi(\br,t)$ is defined by extremization of some functional $\mathsf V$ but here with the sole assumption of unitary symmetry. 
This means that it only depends on $\Psi(\br,t)$ and $\bL(\br)$ through the invariants 
\be\label{invariants} \bmm(t)=\langle\bL|\Psi(\cdot,t)\rangle\, ,\quad s(t,t')=\langle\Psi(\cdot,t)|\Psi(\cdot,t')\rangle\, . \ee
The extremization equation $\delta\mathsf V/\delta\Psi(\br,t)=0$ for such functional $\mathsf V = \mathsf V[\bmm,s]$ reads
\be\label{unitaryEOM} \bL(\br)^\trans\frac{\delta\mathsf V}{\delta\bmm(t)}+\int\d t'\, \frac{\delta\mathsf V}{\delta s(t,t')}\,\Psi(\br,t')=0\, . \ee
Denoting by $K(t,t')$ the inverse of the kernel $\delta \mathsf V/\delta s(t,t')$ (we assume that it exists),
\be\label{Kdef} \int \d t'\, K(t,t')\, \frac{\delta\mathsf V}{\delta s(t',t'')} = \delta(t-t'')\, , \ee
and applying it to the variational equation \eqref{unitaryEOM} then yields once again the sought Eq.~\eqref{unitarypsi} with coefficients
\be\label{unitary-v2} \boldsymbol v(t) = - \int \d t'\, K(t,t')\, \frac{\delta\mathsf V}{\delta\bmm(t')}\, \cvd \ee
This argument is slightly less general than the previous one in that it assumes the inverse model to be defined by a variational principle and requires the existence of the inverse kernel $K(t,t')$.
The result \eqref{unitary-v2} is also less explicit; in particular it was not deduced from it that $\boldsymbol v(t)$ depends on $\bL(\br)$ through the invariants $\langle\bL|\bL^\trans\rangle$.

\section{Expansions and approximations for the spatial leakage model}\label{model-derivation}

In this appendix, we derive the model \eqref{Psi-dec} and its associated approximations (Sec.~\ref{model}) using formal expansions in the state configuration $\Psi_0(\br,t)$. 
We also demonstrate the second key property of the geometric correction scheme stated in Sec.~\ref{gcs-def}.

\begin{center}{\it\small \textbf{Formal expansions}}\end{center}
We begin by introducing the notation
\be\label{psi0split} \Psi_0(\br,t)=\delta(\br-\br_0)\, f(t) + g(\br,t)\, . \ee
Here $f(t)$ represents state activity at location $\br_0$ (as in Sec.~\ref{gcs}) whereas $g(\br,t)$ describes state activity at $\br\neq \br_0$.
The reconstruction $\Psi(\br,t)=(\mathsf W_{\br}\mathsf L\Psi_0)(t)$ of this state (see Eqs.~(\ref{directmodel}, \ref{invop})) can thus be viewed as a functional of $f(t)$ and $g(\br,t)$ that we shall expand in both variables.

To describe the result, it is useful to recall the identity
\be\label{expansionF} F(x+y)=F(x)+F(y)+\mathcal O(xy) \ee
holding for functions with $F(0)=0$ as follows by formally expanding $F(x+y)$ in both $x$ and $y$ around zero.
We shall apply this identity in the following setup, where $F$ is an operator and $x$ and $y$ are field configurations:
\be\label{correspondences} F = \mathsf W_{\br}\mathsf L\, , \quad x = g(\br,t)\, , \quad y = \delta(\br-\br_0)\, f(t)\, .\ee
In general, the condition $F(0)=0$ represents an assumption about the direct and inverse models, albeit a very reasonable one: It means that the reconstruction $\Psi(\br,t)$ of the trivial state configuration $\Psi_0(\br,t)=0$ is itself trivial, $\Psi(\br,t)=0$.
This will be assumed throughout and Eq.~\eqref{expansionF} can thus be applied.

In this setup, the left-hand side $F(x+y)$ coincides with the reconstruction $\Psi(\br,t)$ of the state \eqref{psi0split} since $x+y=\Psi_0$ and $\Psi=F(\Psi_0)$.
The right-hand side of \eqref{expansionF} has three contributions.
The first is $F(x)=\mathsf W_{\br}\mathsf Lg$ and does not depend of state activity $f(t)$ at $\br_0$. This contribution is therefore free from spatial leakage effects emanating from $\br_0$ and corresponds to the field $\Phi(\br,t)$ appearing in Eq.~\eqref{Psi-dec},
\be\label{Phi-g} \Phi(\br,t)=(\mathsf W_{\br}\mathsf L g)(t)\, . \ee
The second is the reconstruction $F(y)$ of the pointlike state $y$ and the definition of the resolving operator (Sec.~\ref{prelim}) shows that $F(y)=\mathsf R_{\br,\br_0}f$, see Eqs.~(\ref{pointlikepsi}, \ref{resop}).
This term represents the \emph{linear} contribution (independent of $g(\br,t)$) of spatial leakage from $\br_0$ and corresponds to the field $\Lambda(\br,t)$ appearing in Eq.~\eqref{Psi-dec},
\be\label{Lambda-f} \Lambda(\br,t)=(\mathsf R_{\br,\br_0}f)(t)\, . \ee
The third and last contribution gathers all cross-terms $\mathcal O(fg)$ describing the nonlinear part of spatial leakage effects from $\br_0$. 

Combining these remarks and definitions together, we finally obtain
\be\label{expansion2} \Psi(\br,t)=\Phi(\br,t)+\Lambda(\br,t)+\mathcal O(fg)\, .\ee
This looks similar to the model \eqref{Psi-dec} except that, in general, (i) Eqs.~(\ref{infer-f}, \ref{Lambda}) do not hold and (ii) the cross-terms are not absent (except when $F$ is linear).
To make contact with the model, we use two independent approximations.

\begin{center}{\it\small \textbf{Approximation (i)}}\end{center}
As said in the main text (Sec.~\ref{model}), the condition necessary for the validity of the estimate \eqref{infer-f} for $f(t)$ is that $g(\br,t)$ does not contribute significantly to the reconstruction at $\br_0$. 
This means that the second term in the right-hand side of Eq.~\eqref{expansion2} must dominate when $\br=\br_0$, i.e.,
\be\label{approx-i} \Psi(\br_0,t) \approx \Lambda(\br_0,t)=(\mathsf R_{\br_0,\br_0}f)(t)\, . \ee
The approximate validity of Eq.~\eqref{infer-f} directly follows from the invertibility of $\mathsf R_{\br_0,\br_0}$.
Plugging the estimate \eqref{infer-f} into the definition \eqref{Lambda-f} yields Eq.~\eqref{Lambda}.

\begin{center}{\it\small \textbf{Approximation (ii)}}\end{center}
We now consider Eq.~\eqref{expansion2} for $\br\neq\br_0$ and make the assumption that the cross-terms $\mathcal O(fg)$ are negligible in front of the two other contributions.
This holds, for instance, in the limit of small field configurations where $\Psi_0(\br,t)=\mathcal O(\epsilon)$ with $\epsilon\rightarrow 0$ (i.e., both $f(t)$ and $g(\br,t)$ are $\mathcal O(\epsilon)$).
Indeed, in this case $\Phi(\br,t)$ and $\Lambda(\br,t)$ equal $\mathcal O(\epsilon)$, too (as $F=\mathsf W_{\br}\mathsf L$ satisfies $F(0)=0$), whereas the cross-terms equal $\mathcal O(\epsilon^2)$.
More generally, this approximation means that we only keep the linear part of spatial leakage effects from $\br_0$ and neglect their nonlinear contribution. 

\begin{center}{\it\small \textbf{Property (ii) of Section \ref{gcs-def}}}\end{center}
A key property for the geometric correction scheme is that, in the limit where the two above approximations hold exactly and in the absence of dynamical interaction between state activity $f(t)$ at $\br_0$ and $g(\br,t)$, no connectivity estimation can disclose a coupling between $\Psi(\br_0,t)$ and $\Phi(\br,t)$.
This statement can be proven directly in the setup developed in this appendix. First, Eq.~\eqref{Phi-g} shows that $\Phi(\br,t)$ depends on the field $g(\br,t)$ but not on the variable $f(t)$, and is thus temporally uncoupled to $f(t)$ by assumption.
On the other hand, Eqs.~\eqref{approx-i} and \eqref{infer-f} show that $f(t)$ is merely a transformed version of $\Psi(\br_0,t)$.
We conclude that $\Phi(\br,t)$ is temporally independent of $\Psi(\br_0,t)$, as claimed.

\section{Linear instantaneous spatial leakage model and relation to the regression model}\label{liinst-derivations}

This appendix fills in details for the quantitative analysis of Sec.~\ref{compare}. 
We first describe an example of Eq.~\eqref{Lambda-liinst} and then derive the relations \eqref{krelation} and \eqref{basicrelation}.

\begin{center}{\it\small \textbf{An example}}\end{center}
We begin by exhibiting a framework in which spatial leakage effects are shown to be linear and instantaneous, see Eq.~\eqref{Lambda-liinst}. 
Quite naturally, we consider direct and inverse models that are both linear and instantaneous. 
Specifically, let us use the direct operator \eqref{directop} and an inverse operator of the form
\be\label{liinst-invop} (\mathsf W_{\br}\bm)(t)=\bW(\br,t)^\trans\bm(t)\, . \ee
This can be viewed as a spatial filter characterized by weights $\bW(\br,t)\in\mathbb R^m$.
The evaluation of the spatial leakage model \eqref{Lambda} requires us to know the resolving operator \eqref{resop}. 
It is obtained by starting from the field configuration \eqref{pointlikepsi}, computing the data predicted by the direct model (\ref{directmodel}, \ref{directop}), i.e., $\bm(t)=\bL(\br_0) f(t)$, and reconstructing it with the inverse operator \eqref{liinst-invop}. 
This leads to 
\be\label{liinst-resop} (\mathsf R_{\br,\br_0}f)(t)=\bW(\br,t)^\trans\bL(\br_0)\, f(t)\, \ee
and thus $ (\mathsf R_{\br_0,\br_0}^{-1}f)(t)=f(t)/\bW(\br_0,t)^\trans\bL(\br_0)$.
Plugging these results into Eq.~\eqref{Lambda} shows that the spatial leakage model is indeed restricted to the form \eqref{Lambda-liinst} with the specific coefficient
\be k(\br,\br_0,t)=\bW(\br,t)^\trans\bL(\br_0)\,/\,\bW(\br_0,t)^\trans\bL(\br_0)\, .\ee
This quantifies the state-independent coupling representing the spatial leakage effect from $\br_0$ to $\br$ at time $t$.

\begin{center}{\it\small \textbf{Relating the coupling constants}}\end{center}
We now derive the relation \eqref{krelation}. 
Trivially rewriting Eq.~\eqref{gcs-field} as
\begin{equation*}\label{gcs-dec} \Psi(\br,t)=\Lambda(\br,t)+\Phi_\gcs(\br,t)\end{equation*}
and plugging this into Eq.~\eqref{regresscoeff}, we find that the regression coefficient $k_\bot(\br,\br_0,t)$ equals to
$$ \mathsf{Re}\big[\Lambda(\br,t)/\Psi(\br_0,t)\big]+\mathsf{Re}\big[\Phi_\gcs(\br,t)/\Psi(\br_0,t)\big]\, . $$
%
The first term estimates the spurious linear instantaneous interaction between seed activity $\Psi(\br_0,t)$ and the spatial leakage model $\Lambda(\br,t)$ and reduces to the coupling coefficient $k(\br,\br_0,t)$ because of Eq.~\eqref{Lambda-liinst}. 
We thus recover the sought Eq.~\eqref{krelation}.

\begin{center}{\it\small \textbf{Relating the corrected fields}}\end{center}
We finally demonstrate Eq.~\eqref{basicrelation} from Eq.~\eqref{krelation}. 
Plugging the latter into Eq.~\eqref{Lambda-orth} shows that
\be \Lambda_\bot(\br,t) = \Lambda(\br,t) + \mathsf{Re}\bigg[\frac{\Phi_\gcs(\br,t)}{\Psi(\br_0,t)}\bigg]\, \Psi(\br_0,t)\, . \ee
Subtracting this from the uncorrected reconstruction $\Psi(\br,t)$ and using the definitions \eqref{Psi-dec-orth} and \eqref{gcs-field} for the corrected ones then yields
\be \Phi_\bot(\br,t) = \Phi_\gcs(\br,t) - \mathsf{Re}\bigg[\frac{\Phi_\gcs(\br,t)}{\Psi(\br_0,t)}\bigg]\, \Psi(\br_0,t) \, .\ee
This is equivalent to the sought Eq.~\eqref{basicrelation}.

\section{Amplitude correlation for a model of coherent Gaussian signals}\label{hippmodelderivation}

In this appendix, we consider the model of coherent Gaussian signals introduced in Ref.~\cite{Hipp2012NatNeurosci} and derive analytically an expression for the true and orthogonalized amplitude correlations.

\begin{figure}
\includegraphics[width=8.5cm]{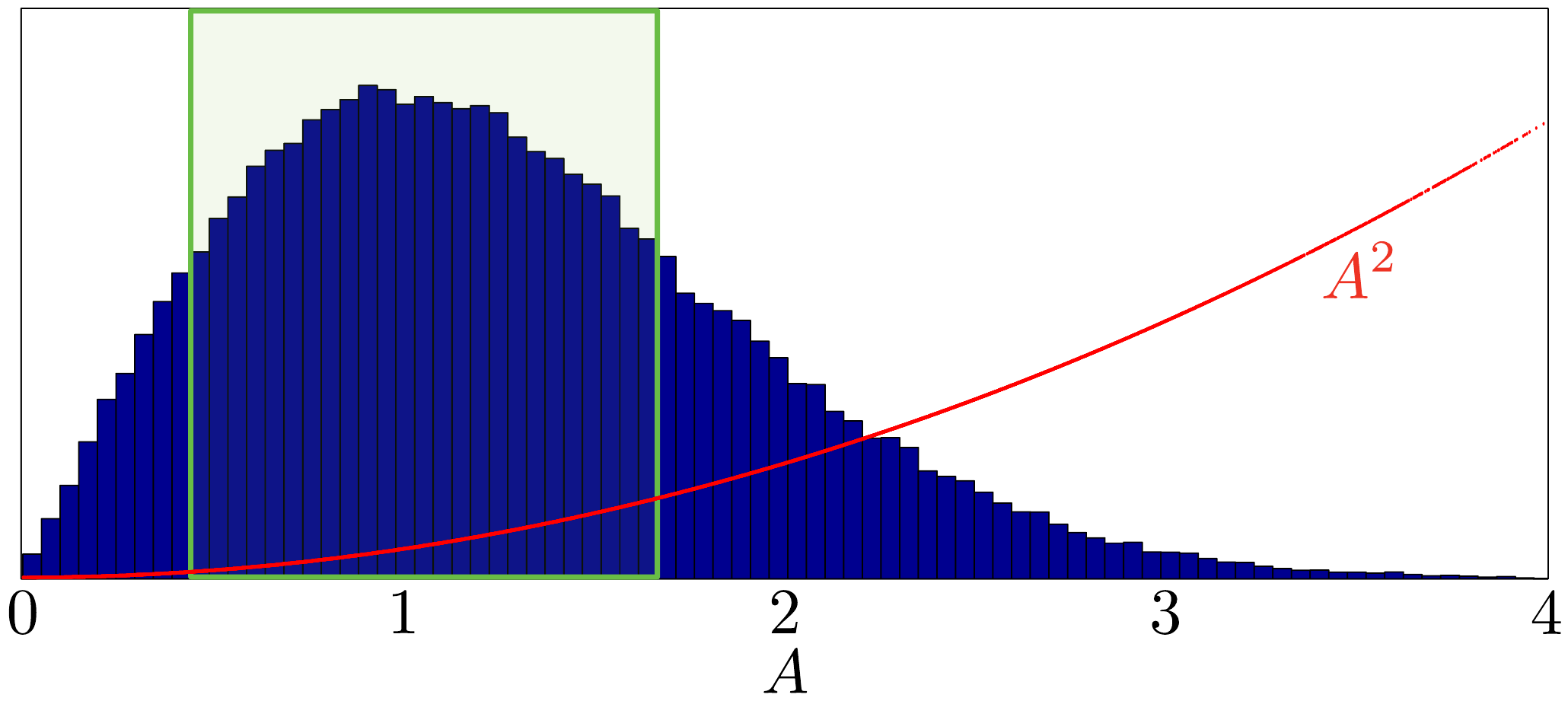}
\caption{\label{fig3} 
\emph{Linear approximation of squared amplitude.} The Rayleigh distribution of $A(t)$ with maximum reached at value $\alpha=1$ is shown together with the function $A^2$. The Taylor expansion $A^2\approx2\alpha A-\alpha^2$ around $A=\alpha$ is seen to be a good approximation in the $\alpha\pm\sigma_A$ interval (green shaded region), which contains about 80\% of samples. Over a larger interval, linear regression still yields a reasonable approximation.} 
\end{figure}

\begin{center}{\it\small \textbf{Model definition}}\end{center}
The model assumes that the spatial-leakage-free reconstruction $\Phi_\gcs(\br,t)$ is linearly related to the seed signal $\Psi(\br_0,t)$. Explicitly,
\begin{align} \label{hippmodel0}\Psi(\br_0,t)&= A_0(t) \exp i\theta_0(t)\, , \\ \label{hippmodel} \Phi_\gcs(\br,t)&=A_1(t) \exp{i\theta_1(t)}+\frac{c\, \exp{i\psi}}{\sqrt{1-c^2}}\, \Psi(\br_0,t)\, .\end{align}
The linear coupling is controlled by two parameters: the coherence parameter $0\leq c\leq 1$ quantifies the strength of this coupling, whereas $\psi$ denotes the mean phase lag between $\Psi(\br_0,t)$ and $\Phi_\gcs(\br,t)$. 
In the absence of coupling ($c=0$), the two fields are assumed to be uncorrelated white noises with identical (complex-valued) Gaussian distribution. 
This means that $A_0(t)$ and $A_1(t)$ are distributed according to a Rayleigh distribution (i.e., their fraction of time spent between values $A$ and $A+\d A$ is $-\d \exp(-A^2/2\alpha^2)$ for some $\alpha>0$, see Fig.~\ref{fig3}) and that $\theta_0(t)$ and $\theta_1(t)$ are uniformly distributed on the circle, all four signals being independent of each other.

Our goal is to evaluate the amplitude correlations
\be\label{corrA} \rho_I =  \mathsf{corr}[A_I,A_0]=\frac{\langle A_I A_0\rangle-\langle A_I \rangle\langle A_0\rangle}{\sigma_{A_I}\, \sigma_{A_0}}\ee
between $\Phi_I(\br,t)$ ($I=\gcs$ or $\bot$) and $\Psi(\br_0,t)$. 
In this definition, $\sigma_A$ denotes the standard deviation 
\be\label{stdA} \sigma_A = \sqrt{\langle A^2\rangle-\langle A \rangle^2}\, .\ee
Computing these nonlinear correlations exactly as functions of the model parameters $c$ and $\psi$ is difficult but illustrates the nontrivial effect of orthogonalization on amplitude correlation. 
This computation was performed numerically in Ref.~\cite{Hipp2012NatNeurosci}.
However, analytical derivations are possible using an appropriate approximation.

\newpage
\begin{center}{\it\small \textbf{Approximation scheme}}\end{center} 
Our basic observation is that all amplitudes in the problem have finite variance, e.g., $\sigma_A^2=(2-\pi/2)\alpha^2$ for the Reyleigh distribution.
Therefore, they are effectively supported on a region where the amplitude squared can be reasonably approximated by a linear function, as illustrated in Fig.~\ref{fig3}. 
Now an elementary property of the correlation coefficient \eqref{corrA} is to be invariant under affine transformations, i.e., 
\begin{equation*} \mathsf{corr}[A_I,A_0] = \mathsf{corr}[a A_I+b, a' A_0+b']\end{equation*}
whatever the coefficients $a$, $b$, $a'$, $b'$. 
Using the choice of these coefficients for optimal linearizations $A_I^2\approx aA_I+b$ and $A_0^2\approx a'A_0+b'$, we obtain the approximate formula
\be\label{corrA2} \rho_I \approx \mathsf{corr}[A_I^2,A_0^2]\, .\ee

\begin{figure}
\includegraphics[width=8.5cm]{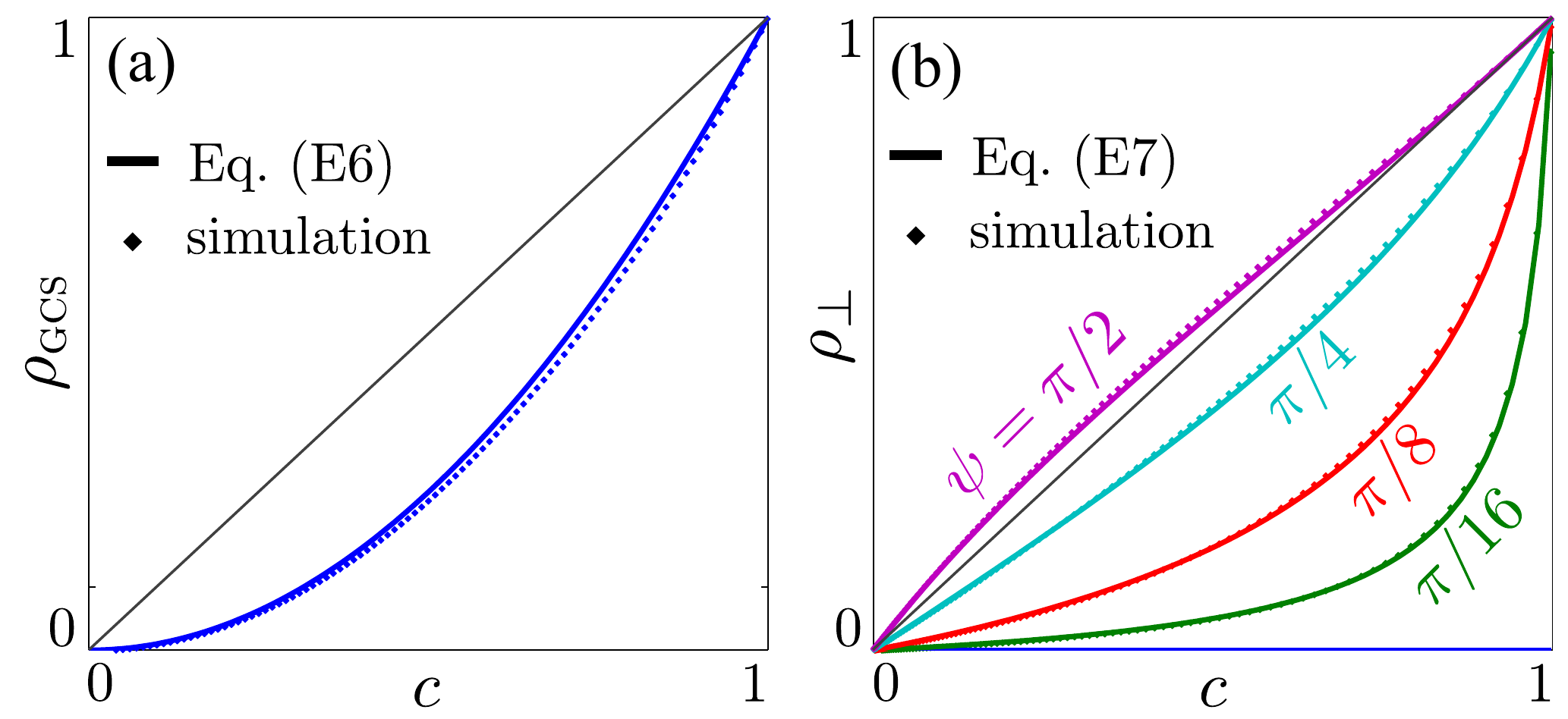}
\caption{\label{fig4} 
\emph{Analytical results versus simulations.} The comparison of Eqs.~\eqref{rhogcs} (a) and \eqref{rhobot} (b) to simulations in which the parameters $0\leq c\leq 1$ and $0\leq \psi\leq \pi/2$ were varied shows very good consistency. For example, fitting a power law $\rho_\gcs=c^k$ on the simulation curve (a) by a log-log linear regression yields $k\approx 2.1$, quite close to the exponent predicted by Eq.~\eqref{rhogcs}.}
\end{figure}

\begin{center}{\it\small \textbf{Solution}}\end{center}
The right-hand side of Eq.~\eqref{corrA2} can be computed analytically in terms of $c$ and $\psi$, as we detail below.
The resulting expressions for $I=\gcs$ and $I=\bot$ are
\begin{align}\label{rhogcs} \rho_\gcs&\approx c^2\, , \\ \label{rhobot} \rho_\bot&\approx\frac{\tilde c^2}{\sqrt{\tilde c^4+2\tilde c^2+1/2}}\, \cvp \end{align}
where we defined the effective coupling constant
\be\label{ctilde} \tilde c=c\sin\psi/\sqrt{1-c^2}\, .\ee
Figure \ref{fig4} shows that these equations based on an approximation agree with numerical simulations.

Eliminating $c$ for $\rho_\gcs$ and plugging into Eq.~\eqref{rhobot} illustrates the nontrivial phase-dependent relation between orthogonalized and true amplitude correlations.
Depending on the values of $c$ and $\psi$, we observe the underestimation and overestimation effects described in Sec.~\ref{amplitude}. 
For example, the case where $\psi=0$ (no phase lag) underestimates amplitude correlation, $\rho_\bot=0$ independently of the coupling strength $c$ or $\rho_\gcs$. (This is because then $\Psi(\br_0,t)$ and $\Phi_\gcs(\br,t)$ are synchronized and their linear instantaneous interaction is completely eliminated by orthogonalization.)
An example of overestimation is given by $\psi=\pi/2$, for then $\rho_\gcs < c < \rho_\bot$ (see Fig.~\ref{fig4}), although it is less drastic than the example \eqref{drasticex}.

\begin{center}{\it\small \textbf{Derivation of Eq.~\eqref{rhogcs}}}\end{center}
To begin our computation of the right-hand side of Eq.~\eqref{corrA2} with $I=\gcs$, we demonstrate that the moments of $A_\gcs(t)=|\Phi_\gcs(\br,t)|$ are given by
\be\label{Agcs-mom} \langle A_\gcs^n\rangle = \frac{\langle A_0^n\rangle}{(1-c^2)^{n/2}}\, \cvp\ee
where $n\in\mathbb N$. 
We will actually show the stronger result that $\Phi_\gcs(\br,t)$ has the same distribution as the Gaussian white signal $A_0(t)\exp{i\theta_0(t)}/\sqrt{1-c^2}$.
We note first that $\Phi_\gcs(\br,t)$ must be a gaussian white noise since it is a linear combination of two independent Gaussian white signals $A_k(t)\exp{i\theta_k(t)}$, see Eqs.~(\ref{hippmodel0}, \ref{hippmodel}). 
It only remains to compute its covariance structure and show that
\be\label{gcscov} \langle\Phi_\gcs(\br,\cdot)^2\rangle = 0\, , \quad \langle A_\gcs^2 \rangle = \frac{\langle A_0^2\rangle}{1-c^2}\, \cvp \ee
from which the result would follow.
The first equation is obtained by explicit expansion of the square of Eq.~\eqref{hippmodel} and using the statistical assumptions about $A_k(t)$ and $\theta_k(t)$.
Each of the three resulting terms vanish. 
For example, let us consider the cross-term, which is proportional to $\langle A_0 A_1 \exp i(\theta_0+\theta_1)\rangle$.
The independence assumption implies its factorization into $\langle A_0\rangle\langle A_1\rangle\langle \exp i\theta_0\rangle\langle \exp i\theta_1\rangle$, which further simplifies to $\langle A_0\rangle^2 \langle \exp i\theta_0\rangle^2$ because distributions are the same for $A_0(t)$ and $A_1(t)$ and likewise for $\theta_0(t)$ and $\theta_1(t)$. 
This term finally vanishes since $\langle\exp i \theta_0\rangle=0$.
The second equation in Eq.~\eqref{gcscov} is obtained similarly by expanding $A_\gcs(t)^2=|\Phi_\gcs(\br,t)|^2$. 
Of the four terms, two vanish for the same reason as above whereas the two others yield
$$ \langle A_\gcs^2\rangle = \langle A_1^2\rangle + c^2 \langle A_0^2\rangle / (1-c^2)=\langle A_0^2\rangle / (1-c^2)$$
since again $\langle A_1^2\rangle=\langle A_0^2\rangle$. 

The computation of $\mathsf{corr}[A_\gcs^2,A_0^2]$ is now a straightforward algebraic exercise.
Details involve similar factorizations and equalities of various expectation values and will not be repeated here.
For the numerator of the correlation coefficient (see the explicit definition \eqref{corrA} for squared amplitudes), we obtain
\be\label{covAgcs2A02}\langle A_\gcs^2\, A_0^2\rangle-\langle A_\gcs^2\rangle\langle A_0^2\rangle = c^2\sigma_{A_0^2}^2 / (1-c^2)\, , \ee
and for the denominator
\be\label{sigAgcs2} \sigma_{A_\gcs^2}\sigma_{A_0^2} = \sigma_{A_0^2}^2/(1-c^2)\, . \ee
Taking their ratio finally yields the sought Eq.~\eqref{rhogcs}.

It is noteworthy that the independence of $\rho_\gcs$ in the phase lag $\psi$ does not only hold within the approximation \eqref{corrA2} but is actually exact, meaning that true amplitude correlation only depends on the coupling strength $c$. 
This can be understood by writing the absolute value of \eqref{hippmodel} as
\be A_\gcs(t) = \Big| A_1(t) + \frac{c\, e^{i(\theta_0(t)-\theta_1(t)+\psi)}}{\sqrt{1-c^2}}A_0(t)\Big |\, ,\ee
which shows that phases only appear in amplitudes through the combination $\theta_0(t)-\theta_1(t)+\psi$.
The basic property 
\be\label{uniftsl} \oint \d \theta_0\, \d \theta_1\, f(\theta_0-\theta_1+\psi)=\oint \d \theta_0\, \d \theta_1\, f(\theta_0-\theta_1)\, ,\ee
for functions $f$ on the circle shows that any expectation value, e.g., of the form $\langle A_\gcs^n A_0^m\rangle$, must be independent of $\psi$.

\begin{center}{\it\small \textbf{Derivation of Eq.~\eqref{rhobot}}}\end{center}
We now turn to the right-hand side of Eq.~\eqref{corrA2} with $I=\bot$.
Plugging Eqs.~(\ref{hippmodel0}, \ref{hippmodel}) into Eq.~\eqref{basicrelation} shows that the orthogonalized field $\Phi_\bot(\br,t)$ equals to
\begin{equation*} i\, e^{i\theta_0(t)} \big(A_1(t) \sin(\theta_1(t)-\theta_0(t)) +\tilde c\, A_0(t)\big)\, .\end{equation*}
This equation indicates that the model parameters $c$ and $\psi$ only appear via the combination \eqref{ctilde} after orthogonalization. 
It also suggests that $\Phi_\bot(\br,t)$ is no longer Gaussian, and the moments of $A_\bot(t)=|\Phi_\bot(\br,t)|$ must be computed explicitly. 
Computations are therefore slightly more cumbersome but still straightforward. 
For the numerator of the correlation $\mathsf{corr}[A_\bot^2,A_0^2]$, we find 
\be\label{covAbot2A02} \langle A_\bot^2\, A_0^2\rangle-\langle A_\bot^2\rangle\, \langle A_0^2\rangle=\tilde c^2\, \sigma_{A_0^2}^2\, .\ee
For the denominator $\sigma_{A_\bot^2} \sigma_{A_0^2}$, we also find that
\be\label{varAbot2} \sigma_{A_\bot^2}^2=(\tilde c^4+3/8)\, \sigma_{A_0^2}^2+(2\tilde c^2+1/8)\,\langle A_0^2\rangle^2\, .\ee 
This can be further simplified thanks to the identity $\langle A_0^2\rangle=\sigma_{A_0^2}$ valid for the Reyleigh distribution, which implies
\be\label{varAbot2-bis} \sigma_{A_\bot^2}^2=(\tilde c^4+2\tilde c^2+1/2)\, \sigma_{A_0^2}^2\, .\ee
The sought Eq.~\eqref{rhobot} follows by taking the ratio of Eq.~\eqref{covAbot2A02} and $\sigma_{A_\bot^2} \sigma_{A_0^2}$ with Eq.~\eqref{varAbot2-bis}.


\bibliography{./references}


\end{document}